\documentclass[11pt,leqno]{article}

\usepackage{amsmath,amssymb,amsfonts,amsthm,bbm,mathrsfs,verbatim} 
\usepackage{setspace}
\usepackage{authblk}
\usepackage{caption}
\usepackage{geometry}
\usepackage{graphics} 
\usepackage{graphicx}  
\usepackage{hyperref}
\usepackage{fancyhdr} 
\usepackage{color}
\usepackage{bm}
\usepackage{algorithm2e}
\usepackage{apacite}

\theoremstyle{definition}

\def\Psib{{\bf \Psi}}
\def\R{{\mathbb R}}

\def\E{{\mathbb E}}

\def\btheta{\bm \theta}

\def\Stil{ \widetilde{S} }

\newcommand\blue[1]{\textcolor{blue}{#1}}

\title{Static hedging of freight rate risk in the shipping market under model uncertainty} 
		
\date{}

\author[a,b]{\small G. I. Papayiannis\footnote{Corresponding author: 80 M. Karaoli \& A. Dimitriou, Piraeus, 18534, Greece. E-mail: \blue{\texttt{gpapagiannis@unipi.gr}}}}

\affil[a]{\footnotesize Department of Statistics and Insurance Science, Risk Management and Insurance Laboratory, Univeristy of Piraeus, Piraeus, Greece}
\affil[b]{\footnotesize Stochastic Modeling and Applications Laboratory, Athens University of Economics \& Business, Athens, Greece}

\begin{document}
\maketitle

\begin{abstract}
Freight rate derivatives constitute a very popular financial tool in shipping industry, that allows to the market participants and the individuals operating in the field, to reassure their financial positions against the risk occurred by the volatility of the freight rates. Management of freight risk is of major importance to preserve the viability of shipping operations, especially in periods where shocks appear in the world economy, which introduces uncertainty in the freight rate prices. In practice, the reduction of freight risk is almost exclusively performed by constructing hedging portfolios relying on freight rate options. These portfolios needs to be robust to the market uncertainties, i.e. to choose the portfolio which returns will be as less as it gets affected by the market changes. Especially, at time periods where the future states of the market (even in the short term) are extremely ambiguous, i.e. there are a number of different scenarios that can occur, it is of great importance for the firms to make decisions that remain robust to these uncertainties. In this work, a framework for the robust treatment of model ambiguity in (a) modeling the freight rates dynamics employing the notion of Wasserstein barycenter and (b) in choosing the optimal hedging strategy for freight risk management, is proposed. The capabilities of the proposed method are demonstrated through two numerical experiments: (a) a carefully designed synthetic data study in which the robustness of the method is assessed at different levels of heterogeneity of the prior set, set sizes and hedging horizons, and (b) a real data experiment with case studies from two different trade routes in which the hedging performance of the method at time periods that the shipping market was at shock is assessed with satisfactory results.
\end{abstract}



\section{Introduction}

Seaborn trade has been developed rapidly the last fifty years since the majority of international trade transportations and supply chain activities are performed almost exclusively through the shipping industry. Valuable and necessary commodities for the growth and functionality of the economies like crude oil, oil products, iron ore, coal grain and others, are primarily chartered between ports through cargo ships using the international sea trade routes. Alongside with the increasing activity in the shipping industry concerning worldwide transportations, it came up the need for all parties involved, to constitute a separate market equipped with financial products that will allow them to perform their operations at acceptable levels of financial risk \cite{alizadeh2009shipping, kavussanos2021freight}. The most important source of risk that appears in shipping operations, is the so called freight risk, which concerns the volatility in the freight rate prices, on which mostly depends the cost of freight services. Freight rates could significantly differ across various sea trade routes and time periods, inducing significant operational hazards that need to be taken into account by shipping firms or firms that use freight services in order to remain viable.

The involved parties in seaborne trade operations (typically the shipowners and the charte\-rers) always seek for financial (hedging) strategies that will allow them to control the induced freight risk. This type of risk is mitigated using relevant financial instruments, which in general are referred to as \emph{freight derivatives}, that allow hedging against the risk induced by the changes in the freight rates. Freight derivatives are traded either as freight futures or options at different maturity dates. Freight options are contracts very similar to vanilla options and are written on the difference of the freight rate spot price from a specified strike price. Freight future contracts concern mainly the freight rates for dry bulk carriers and tankers and depending on the cargo different quotations are applied on the rate, e.g. dry bulk futures are contracts quoted in USD per day while tanker freight futures are quoted on USD per ton. Of course contracts like swap futures, container freight swap agreements, container freight derivatives, etc. are also considered as freight derivatives. The role of the market for the freight derivatives, is played by the Forward contract Freight Agreement (FFA) market, which acts in parallel to the physical market (i.e. the physical world where contracts are conducted according to the spot freight rates), allowing for the trade of the aforementioned financial products related to the freight rates. Being a relatively new market, and due to the special nature of the freight service as a commodity (non-storable like electricity), the typical market completeness assumption does not hold in general \cite{adland2021statistical}, since the deals are typically conducted over the counter (OTC) through broker agencies (e.g. Forward Freight Agreement Brokers' Association (FFABA)). FFA market is characterized by very high volatilities, similar to those observed in the electricity markets, and the freight rate prices mainly depend on the relation between demand and supply for freight services. Despite the aforementioned inconsistencies and singularities, FFA market traded products are the most preferable (and appropriate) hedging instruments for the operating parties in the shipping sector, being by construction strongly related to the physical market's assets (spot freight rates). Although FFA market offers the financial mechanism for managing freight risk, the uncertainty that characterizes several aspects of the freight risk management task needs to be carefully assessed and treated.

Modelling and predicting freight rate prices is a very active scientific area, where several directions in successfully forecasting different phenomena displayed by the freight rates when the stationarity assumption holds have been discussed in the relevant literature (see e.g. \cite{batchelor2007forecasting, chen2012forecasting, katris2021time, munim2021forecasting}). A single model is in general not able to sufficiently capture all aspects of the random behaviour displayed by the freight rates as long as different characteristics may appear depending on the market conditions. Several approaches have been examined so far, capturing various characteristics displayed occasionally by the freight rates dynamics, e.g. seasonality patterns \cite{kavussanos2001seasonality}, mean-reversion effects \cite{benth2015stochastic}, rapid and steep changes in prices which could be interpreted either as jumps \cite{kyriakou2017freight} or potential regime switches \cite{alizadeh2015regime}, etc.

The task of hedging the incurred risk from freight rate prices is a problem of major importance for the shipping industry. Unavoidably, this financial operation strongly depends on the model used for describing the evolution of the freight rates. Based on the aforementioned modelling considerations and others, research contributions of the field of hedging the freight risk has importantly increased the latter years. In particular, a numerous approaches have been discussed in the relevant literature either enriching the classical minimum variance principle (see e.g. \cite{kavussanos2006shipping}) with more sophisticated models, or introducing new criteria in order to reduce the magnitude of error. To name just a few, the standard and widely used delta method has been coupled with: (a) mixed autoregressive-GARCH models \cite{kavussanos2004market} to better capture volatility changes, (b) mixed regime switcing - GARCH models \cite{alizadeh2015regime} to allow for the inclusion of different volatility patterns, and (c) with factor modelling approaches for FFA and spot prices dynamics \cite{prokopczuk2011pricing} in an attempt to better calibrate the stochasticity observed in the prices. More recentlly, approaches that deviate from the classical minimum variance principle have been introduced, e.g. employing time varying copula models combined  with the minimum VaR-hedge ratio principle \cite{shi2017time}, quantile regression approaches \cite{gu2020quantile}, probabilistic approaches in selecting the optimal hedge ratio employing quantile-based criteria instead of mean estimates \cite{sel2022hedging}, and others.

Besides the important contributions in the modelling perspective of the freight rate dynamics and in adopting different type of models and criteria for the risk hedging task, the issue of model uncertainty has not yet been considered. In fact, there exist only some attempts in calibrating behavioural aspects of the market agents acting under uncertainty (see. e.g. \cite{ishizaka2018evaluation} and references therein). Recent evidence indicates that the shipping market is highly affected by disturbances in the world economy, e.g. Global Financial Crisis, political and economical crisis to major countries for the shipping sector, COVID pandemics, etc. As a result, in such conditions the market abandons its stable state and  ambiguity grows concerning the most plausible model to describe even the near future. Information collected by market agents and experts (panellists) is used to derive estimates for the forthcoming situation, however depending on the states of the market and other factors, the received information could be quite heterogeneous and even missleading. Under such circumstances, the major priority of the risk manager in charge is to properly discount the market information input and derive financial decisions/strategies that will remain robust to the scenario that will actually occur. Robustness property in modern decision making was formally introduced in a quantitatively framework about two decades ago in \cite{hansen2001robust, hansen2011robustness, maccheroni2006ambiguity}. The meaning of remaining robust to uncertainty when making a financial decision, refers to the low degree of sensitivity of the deduced decision with respect to the scenario that actually materializes. Under the framework of model uncertainty, the robustness of the decision making process is twofold and refers to (a) the optimal action/decision/strategy to be chosen, and (b) aggregation/combination of the provided information (models) on which strongly depends the optimal strategy to be selected.

In this paper, a framework that handles model uncertainty issues that appear in shipping markets is proposed, bringing to the first place the appropriate design of effective and robust to model uncertainty hedging schemes for managing freight risk. The discussed approach is based on the aggregation of various information sources, identified by certain probability models, through barycentric approaches that have been successfully introduced in decision theory \cite{petracou2022decision}, in economics \cite{papayiannis2022robust,koundouri2024consensus} and in financial risk quantification \cite{papayiannis2018convex, papayiannis2023framework}. The presented framework is directly implemented in designing the decision rule for selecting the optimal hedging strategy, that will remain robust to model uncertainty, for reducing the exposure of a position to freight risk. As a first attempt in this direction, the problem is considered in the static framework, i.e. the hedging policy is selected at a certain time instant, however the discussed approach is extendable to the multi-period or the dynamic setting. The paper is organized in the following way: in Section 2 some preliminaries on shipping freight market and its derivatives are provided along with standard modelling approaches for freight rate dynamics and related model uncertainty issues, while in Section 3 are presented the main methodological contributions of the paper, deriving and characterizing robust optimal hedging strategies within the model uncertainty setting. Finally, in Section 4 the proposed method capabilitieas are assessed through two numerical experiments: (a) a detailed simulation study employing standard models that are used in describing freight rate dynamics, illustrating the capabilities of the proposed methodology and testing its sensitivity with respect to varying levels of information heterogeneity, and (b) implementing the method to real data involving two different types of sea vessels (Panamax and Capesize) at specfic periods of the last twenty years that significant shocks occurred in the shipping market.

\begin{table}[ht!]\scriptsize
	\centering
	\begin{tabular}{l|l}
		\hline\hline
		{\bf Notation} & {\bf Explanation}\\
		\hline
		$\E_{Q}[\cdot]$ & expectation with respect to the probability measure $Q$\\
		$\mathcal{F}_t$ & $\sigma$-algebra containing the available information up to time instant $t$\\
		$(x)_{+}$ & positive part of $x$, i.e. $(x)_{+} = \max(0,x)$\\
		$\| x \|_2$ & Euclidean norm of $x$\\
		$\mathcal{P}(\Omega)$ & the space of probability models with support in the set $\Omega$\\
		$\Delta^{n-1}$ & the $n$-dimensional unit simplex, i.e. $x\in\R^n$ such that $\sum_{i=1}^n x_i=1$ and $x_i \geq 0$ for all $i$\\
		$\mathbb{P}(d)$ & the set of symmetric and positive definite matrices of dimension $d \times d$\\
		\hline\hline
	\end{tabular}
	\caption{Explanation of notations used throughout the text}
\end{table}

\section{Freight Derivatives and the Framework of Model Uncertainty}\label{sec-2}

\subsection{Freight Derivatives Market and Freight Options}
 
In 1985, the Baltic Exchange\footnote{\url{https://balticexchange.com/en/index.html}}, supported by shipping market brokers, has undertaken a central role in constructing independently prices for the freight service of the ocean-going cargo carrying vessels, the so called freight rates, that could be used as a base to be written futures and option contracts. This was succeeded by collecting information from a group of designated brokers, referred to as the panelists, who are the middle persons between charterers and shipowners being aware of the prevailing market freight rates in individual routes. The Baltic Exchange started collecting these freight rates, calculated weighted averages of them and reported them in the form of indices for freight services. These indices are published on a daily basis, for each sector of the shipping industry, i.e. dry-bulk, tanker and container sector, for the major trade routes that the shipping operations take place. Over the years, more appropriate indices are constructed and revised for the better and more accurate monitoring of the freight rate prices. For instance, in the dry-bulk sector, the Baltic Exchange Dry Bulk Index (BDI) was launched in 1985, the Baltic Panamax Index (BPI) was launched in 1998, the Baltic Capesize Index (BCI) appeared in 1999 and was revised in May 2014 (BCI 2014) and the Baltic Supramax Index (BSI) was launched in 2005 (for more details on the subject please see the excellent references \cite{alizadeh2009shipping} and \cite{kavussanos2021freight}). The individual dry-bulk and tanker routes or baskets of routes of the Baltic Exchange Freight Indices serve as the underlying assets of freight derivative contracts. 

Forward freight agreements (FFAs) introduced in early 90's as over-the-counter (OTC) derivatives contracts. Typically, such agreements constitute private contracts between a seller and a buyer for settling a freight rate, for a specified cargo or type of vessel, for either one or a combination of the major trade routes \cite{alexandridis2018survey}. Freight rate options are path-dependent contingent claims, and in particular, Asian type derivatives written on the spot freight rates which are non-traded in the market, depending on a number of predefined dates through their arithmetic average, to avoid market manipulation phenomena especially close to the maturity of the options. An FFA is a cash-settled financial contract that provides to the owner of the contract the difference between the average of the spot freight rate prices $S(t)$ at a number of predefined time instants (dates) $T_1,T_2,...,T_N$ and the future price $F(t,T_1,T_N)$ multiplied by a factor $D$ (cargo size in tonnes or number of days for the charter). The value of an FFA can be calculated by discounting the cash flow received at the maturity time instant $T_N$ taking conditional expectation under the pricing measure $Q$ (however, according to \cite{adland2021statistical} there exist statistical arbitrage profits in the freight option markets, suggesting a degree of market inefficiency). Since there is no cost in entering to an FFA, the expectation can be taken equal to zero, i.e.
\begin{equation}
\E_Q\left[ e^{-r(T_N - t)} D \left( \frac{1}{N}\sum_{i=1}^N S(T_i) - F(t, T_1, T_N) \right) \mid \mathcal{F}_t \right] = 0
\end{equation} 
where $r>0$ denotes the risk-free interest rate and $\mathcal{F}_t$ the information for the spot prices up to time $t$. Solving the above with respect to $F(t, T_1, T_N)$ we obtain the representation
\begin{equation}\label{FFA-eq}
F(t, T_1, T_N) = \frac{1}{N} \sum_{i=1}^N \E_Q[ S(T_i) | \mathcal{F}_t]
\end{equation}
where $Q$ denotes the equivalent martingale measure satisfying the market completeness assumption. The latter relation is useful for determining the dynamics of the $F(t,T_1,T_N)$ by specifying the stochastic model which describes the spot freight rates dynamics. 

An Asian option of this type can be interpreted as a European option on the forward contract value $F(t,T_1,T_N)$. For any strike price $K$ and maturity time $T=T_N$, we obtain either a call or a put option. A call option on the FFA contract referred to as a caplet, provides freight rate protection for the buyer (typically a charterer) above a predetermined level (the cap rate) and its price is calculated by the formula
\begin{equation}
C(t,T) = e^{-r(T-t)} D \E_Q\left[ ( F(T, T_1, T) - K )^+ | \mathcal{F}_t \right].
\end{equation}
A put option on the FFA contract referred to as a floorlet, guarantees downside protection on the freight rates for the buyer (typically the shipowner) at a predetermined level (the floor rate) and its value is derived by the formula
\begin{equation}
P(t,T) = e^{-r(T-t)} D \E_Q\left[ (  K - F(T, T_1, T)  )^+ | \mathcal{F}_t \right].
\end{equation}
Unfortunately, closed-form pricing formulas for these options cannot be in general derived, even in the typical setting of a geometric Brownian motion model for the spot dynamics, due to the arithmetic average term. However, in \cite{koekebakker2007pricing} an analytic expression is derived under some assumptions leading to a log-normal approximation.

\subsection{Modelling freight rate dynamics}

Several approaches have been proposed in the literature so far for modelling commodities and in particular freight rate prices. Here we briefly discuss some well celebrated models that are employed in practice. The first model appeared in the literature for modelling commodities prices is the famous Black's model \cite{black1976pricing}. According to this approach, the spot freight-rate prices $S(t)$ are described by the classical geometric Bronwian motion (GBM) model
\begin{equation}
dS(t) = \mu S(t) dt + \sigma S(t) dW(t)
\end{equation}
where $\mu, \sigma$ denote the drift and the volatility coefficients while $W(t)$ is the standard Brownian motion process. In this perspective, the price of the modelled commodity (in our case the rate for the freight service) is treated as any other risky asset in the market. Although for certain time periods this model could be adequate to describe the evolution of the freight rate prices, it cannot efficiently capture typical phenomena that occur in this high volatile market, e.g. rapid changes, seasonality trends, etc. A second attempt in modelling commodity prices appeared in \cite{schwartz1997stochastic}, where the mean reversion model (MR) was employed to capture mean-reversion effects. In this case, freight rate prices are described by the model
\begin{equation}\label{OU}
dS(t) = \alpha ( \mu - S(t) )dt + \sigma dW(t)
\end{equation} 
where $\mu$ denotes the long-term average of the process, $\alpha>0$ denotes the mean-reversion rate towards the mean level $\mu$, $\sigma$ being the volatility parameter and $W(t)$ the typical Brownian motion. This approach has been proved very successful in capturing the typical mean-reversion behavior that appears in many commodities. A more flexible two-factor model framework has also appeared in the literature (both GBM and MR model are cases of one-factor models) where the freight rate prices (or the log-prices) are modelled by the sum of two separate processes with possibly different characteristics, e.g. a GBM model and a MR model, or combinations of them or other stochastic models (see e.g. \cite{prokopczuk2011pricing} and references therein). Even more interesting approaches have been examined the latter years where jump diffusion processes are employed to capture the phenomena of rapid changes in freight rate prices. Such models were investigated either using as a basis GBM models or MR models, enriched with jump terms, providing a very flexible framework in capturing different characteristics of the market. Some recent attempts in this direction with applications in pricing freight rate options can be found in \cite{gomez2021including, kyriakou2017freight}.

\subsection{The issue of model uncertainty in freight markets}

FFA market presents high and many times extreme volatilities in freight rates since prices are directly affected by the demand and supply levels for freight services that are observed in the market (similarly to the behaviour of electricity markets). Analysis of historical data establishes the effect of seasonality in the freight rates which this effect can be captured by employing GARCH-type models \cite{kavussanos2001seasonality}. However, there are several characteristics that are occasionally displayed by the freight rate dynamics for which there is not a single universally acceptable model that can capture them. 

For example, a model with mean-reversion terms should generally be acceptable for the description of the freight service prices. However, depending on the world economy situation, it could appear large periods where the mean reversion behaviour is not observed at all and employing such a term would be misleading and disastrous for financial decisions. A recent example is the world economic crisis in 2008, where all markets, including the shipping market, were on the downside for a very large time period \cite{samitas2010hedging}. More recently, the COVID-19 pandemics significantly affected the market operations \cite{kamal2021stock, notteboom2021disruptions}. Such events seriously disturb the economic activity at all levels and introduce long-term effects in the market which may enter different states for a large period of time with much different patterns than the usual ones to be observed \cite{ishizaka2018evaluation}. Under such circumstances, due to the high level of ambiguity, the operational risk is highly increased, and it is not clear which is the most appropriate model to be used for discounting the decisions. As part of the confusion that governs the market, incoming information from market sources concerning the forthcoming situation may be quite divergent leading to the classical model uncertainty setting, where there is a number of provided models (predictions) however it is not clear which one should be trusted. Obviously, models that have been calibrated using historical data when market was in a steady state, are practically useless to discount decisions when the market is in shock. Even in case where certainty for the occurrence of specific phenomena in the near future that will affect the price dynamics exists (e.g. certain drift effects, rapid fluctuations like downward or upward price shocks), there is ambiguity concerning their intensity. For instance, consider the case that upward jumps are expected to the freight rates due to a rapid increase in the demand of freight service at a certain trade route \cite{nomikos2022disentangling}. In such a case, the magnitude of the jump in the freight service prices cannot be precisely estimated and there may be several scenarios on that, depending on the evolution of the demand for the freight service in the route.

All the aforementioned cases need a special treatment during the decision making process. Practically, a shipping firm that desires to hedge the induced freight risk from its shipping operations under such conditions, faces the problem of model uncertainty, i.e. several scenarios could happen in the near future but it is extremely difficult, if not impossible, to distinguish which scenario will actually occur. In such a case, the various possible realizations (scenarios) must be handled in a robust manner, to reduce the sensitivity of the hedging decision that is chosen with respect to the scenario that actually materializes. A framework that is appropriate for handling model uncertainty in problems of this type has been proposed in \cite{papayiannis2018convex, papayiannis2023framework} and \cite{koundouri2024consensus,petracou2022decision} under the perspectives of financial risk management and group decision making, respectively. In the aforementioned approaches, various scenarios or opinions are identified by certain probability models, and the notion of the Wasserstein barycenter (roughly the sense of median in the space of probability models) is employed to estimate an aggregate model that robustly represents the received information and which is then used to derive a robust decision rule. In what follows, these ideas are carefully employed under the context of hedging freight risk under model ambiguity.

\section{Static hedging of freight rate risk under model uncertainty}\label{sec-3}

Although dynamic hedging has a long history in financial risk management, there is an increasing trend among academics and industry practitioners towards abandoning it and turning more attention on static approaches \cite{carr1998static, carr2014static} which are easier and cheaper to implement and in some cases have been proved a better option than the commonly used delta hedging. Static hedging is suggested in hedging of exotic type derivatives that are path-dependent \cite{carr1998static, leung2016optimal, kirkby2019static}, like the ones used in the shipping industry (e.g. FFA-options), and in many cases is possibly the only option due to the nature of the market (e.g. market incompleteness issues like the FFA market which does not allow for dynamic hedging). Comparisons in the literature that have been conducted between static and dynamic hedging approaches, do not indicate that the dynamic approach is always the better one (see e.g. \cite{tompkins2002static, adland2020hedging}). Static hedging can be either exact or approximate, the latter term referring to the construction of a static portfolio approaching a required target, e.g. the payoff of a financial obligation or a contingent claim to be hedged as close as possible with the concept of closeness being quantified by an appropriate measure of distance. Morover hybrid-approaches have been also appeared in the literature (see e.g. \cite{boyarchenko2020static} and references therein) and robastification approaches that mitigate potential drawbacks of the static approaches (see e.g. \cite{maruhn2009robust}). In this section, static hedging strategies are derived under the model uncertainty framework for reducing the exposure to freight rate risk for the charterer, while characterizations of the pricing measure are provided under standard modelling considerations.


\subsection{A robust principle for choosing the pricing measure }

Consider the case where an owner of a forward freight contract (FFA) desires to construct a hedging strategy for this contract. Since the value of the contract depends on the condition of the market, the final outcome of this deal may significantly vary, depending on the situation in both markets (physical and FFA market). As a result, the contract owner needs to secure her/his position against to market volatilities by appropriately selecting her/his hedging strategy. Due to the discussed limitations in the shipping freight rates market, the static hedging approach seems to be the most prominent choice, especially when the time period to be covered is short. Otherwise, a static hedging approach could be also applied in a multi-stage framework (i.e. approximating delta hedging through multi-stage static hedging). The main issue with freight options is their dependence on underlying assets (the spot freight rates) which are non-tradable, therefore many kind of uncertainties arise. For example, following the discussion in Section \ref{sec-2}, when the market is not on equilibrium or is perturbed by certain events, it is not straightforward how to choose a plausible model for the spot price and the FFA dynamics. Different choices may arise depending on the situation, e.g. a period with typical market situation may require a model with mean-reversion effects, a period of higher volatility may be best described employing a more volatile model like geometric Brownian motion, inside information by various market agents/sources might also be taken into account, etc. Therefore, an appropriate choice of model should be robust to these uncertainties, i.e. a robust model should lead to a decision that whatever scenario is realized, the optimal derived decision should be scenario-insensitive.

Let us introduce a framework to treat model uncertainty through a simple but very appropria\-te example for the typical situation in the freight market. Assume that the spot freight rates of a certain cargo vessel (non-traded asset) is denoted by $S(t)$ and a related market index for the freight rates at the specific trade route that the vessel operates is denoted by $\Stil(t)$. Given that the market risk-free rate is $r>0$, the processes $S(t),\Stil(t)$ are either assumed to follow (a) a two-dimensional Black's model (GBM) described by the dynamics
\begin{eqnarray}\label{gbm-mod}
\begin{array}{c}
dS(t)    = \mu S(t) dt + \sigma S(t) dW(t)\\
d\Stil(t) = \tilde{\mu} \Stil(t) dt + \tilde{\sigma} \Stil (t) d\widetilde{W}(t)\\
\E_Q[dW(t)d\widetilde{W}(t)] = \rho dt
\end{array}
\end{eqnarray}
where $\mu, \tilde{\mu}$ denote the respective drifts, $\sigma,\tilde{\sigma}$ the respective volatilities, and $W(t),\widetilde{W}(t)$ the respective correlated Brownian motions (with $\rho \in (-1,1)$ denoting the correlation coefficient), or (b) a two-dimensional Schwarz's model (MR) given by the dynamics
\begin{eqnarray}\label{OU-mod}
\begin{array}{c}
dS(t)    = \alpha (\mu - S(t)) dt + \sigma dW(t)\\
d\Stil(t) = \tilde{\alpha}(\tilde{\mu} - \Stil(t)) dt + \tilde{\sigma} d\widetilde{W}(t)\\
\E_Q[dW(t)d\widetilde{W}(t)] = \rho dt
\end{array}
\end{eqnarray}
where $\mu, \tilde{\mu}$ denote the respective long-term average rates, $\alpha, \tilde{\alpha}$ the mean-reversion intensity parameters, $\sigma,\tilde{\sigma}$ the respective volatilities and $W(t),\widetilde{W}(t)$ as above. Let us denote by $Q$ the probability measure (model) that describes the joint dynamics of $S,\Stil$. An investor holds a FFA contract (let us say a charterer) depending on the spot prices $S(t)$ and wishes to cover her/his exposure through a hedging portfolio depending on the index $\Stil(t)$ (e.g. a portfolio of derivatives and forward contracts written on $\Stil(t)$). Since, the spot freight rates $S(t)$ is a non-tradable asset, and due to the lack of sufficient statistical data, various uncertainties are introduced as to the exact values of the parameters of the true model. The investor to counter the uncertainty concerning the true (but unknown) model $Q$, is provided with a collection of $m$ models 
$$\mathcal{M} = \{Q_1, Q_2, ..., Q_m\}$$ by her/his network of market agents/sources. Each model possibly contains different fragments of reality and therefore the investor cannot fully allocate her/his trust only to one of these models. The issue arising here, is how to efficiently aggregate the available information into a single model which will be used to robustly discount the outcomes of the financial decisions to be made.

A robust and efficient approach in combining different beliefs into a single aggregate model has been proposed in \cite{papayiannis2018convex} and \cite{petracou2022decision} where the notion of the Wasserstein barycenter (derived from the generalized mean sense offered by the notion of Fr\'echet mean \cite{frechet1948elements}) is combined within the frameworks of convex risk measures and expected utility to properly discount decisions under model ambiguity. The Wasserstein distance is a proper metric in the space of probability models (see for technical details \cite{santambrogio2015optimal, villani2008optimal}) and between two probability models $Q_1, Q_2 \in \mathcal{P}(\Omega)$ is computed (according to Kantorovich formulation) as the minimal cost of the problem
\begin{equation}
W_2^2(Q_1, Q_2) = \min_{\Gamma \in \Pi(Q_1, Q_2)} \int_{\Omega \times \Omega} \|x-y\|_2^2 d\Gamma(x,y) 
\end{equation} 
where $\Pi(Q_1, Q_2)$ is the set of transport plans with marginals the measures $Q_1, Q_2 \in \mathcal{P}(\Omega)$. Since Wasserstein distance is fully compatible with the geometry of the space of probability models, constitutes a very effective instrument for the quantification of discrepancies between various opinions which can be identified by certain probability models. On the examined framework where a prior set with a multitude of models is provided, the notion of Wasserstein barycenter is employed (see e.g. \cite{agueh2011barycenters} for technical discussion), i.e. an appropriate analog of the mean sense in the space of probability models (opinions), to derive a single aggregate model comprising all the available information from the set $\mathcal{M}$. Given a weighting vector $w \in \Delta^{m-1}$, which can be considered as each model's contribution to the final aggregate, the Wasserstein barycenter is defined as the minimizer of the respective Fr\'echet variance
\begin{equation}\label{wbar}
Q_*(w) = \arg\min_{Q \in \mathcal{P}(\Omega)} \sum_{i=1}^m w_i W_2^2(Q,Q_i).
\end{equation} 
In general, problem \eqref{wbar} does not admit a closed-form solution, however for the case that all models in $\mathcal{M}$ are members of a Location-Scatter family (e.g. Gaussian), which is the case in the discussed setting with GBM or MR models, the barycenter of the set $\mathcal{M}$ can be characterized through its parameters in a semi-closed form. In particular, the location of $Q_*(w)$ is represented as the weighting average 
\begin{equation}\label{wloc}
{\bf m}_B(w) = \sum_{i=1}^m w_i {\bf m}_i
\end{equation} 
with ${\bf m}_i$, $i=1,2,...,m$ denoting the location parameters of models $Q_1,Q_2,...,Q_m$, while the dispersion characteristics of $Q_*(w)$ are represented by the covariance matrix $C_B$ which satisfies the matrix equation
\begin{equation}\label{wcov}
C_B = \sum_{i=1}^m w_i \left( C^{1/2}_B C_i C^{1/2}_B \right)^{1/2}
\end{equation}
with $C_i$ for $i=1,2,...,m$ denoting the covariance matrices of models $Q_1,Q_2,...,Q_m$. The matrix $C_B$  can be obtained numerically by the fixed-point scheme  proposed in \cite{alvarez2016fixed}. For the case of non Location-Scatter models, numerical schemes based on the entropic regularization of the main problem to achieve higher convergence rate have been proposed in the literature (see e.g. \cite{cuturi2013sinkhorn,carlier2017convergence, clason2021entropic}).

The barycenter stated in \eqref{wbar} serves as the aggregate model of the set $\mathcal{M}$ depending on the weighting vector $w$. The latter can be realized as a sensitivity parameter chosen by the investor, determining the level of contribution/influence of each provided model in $\mathcal{M}$ to the model that will be actually used to discount the decisions. For instance, if $w_i=0$ for a certain $i$ is set, then this particular model is omitted as unrealistic or untrustworthy. Similarly, if $w_i=1$ is chosen, then the investor fully allocates her/his interest to this particular model and omits the rest. Equal weighting, i.e. $w_1=w_2=...=w_m=1/m$, corresponds to the typical barycenter case where no extra information about the credibility of each model is available. In the case that these models concern different scenarios that could be realized (and they do not represent just different opinions), the choice of the discounting probability measure in this way allows for determining a robust strategy across scenarios avoiding to concentrate to just one of them. In this perspective, $w$ could be understood as the probability (possibly in a subjective view) for each one of the scenarios to occur. Weighting the various scenarios and not concentrating to just one of them, should lead to optimal decisions/strategies which financial output will be less affected by the scenario that actually occurs. This is the essence of robustness, however the choice of weighting vector $w$ and the quality of the information provided by the prior set $\mathcal{M}$ (does it contains the ``true'' scenario?) can act as sensitivity parameters and should be carefully chosen and assessed. However, hybrid schemes combining aversion preferences with data-driven approaches in updating the weighting vector whereas new information batches are available can be developed by applying appropriate scoring rules (see, e.g. \cite{papayiannis2018learning, koundouri2024consensus}). 

Clearly, there is always the chance that no model in $\mathcal{M}$ is realized or the investor does not fully trust none of these models. In that case, the investor may quantify her/his aversion variationally in the spirit of \cite{hansen2001robust, maccheroni2006ambiguity} and the deduced model would be a distorted version of the barycenter model as computed by the set $\mathcal{M}$ (please see \cite{papayiannis2018convex} and the related results therein).  The level of distortion depends on the level of reliability that the investor allocates to the provided information with the special cases corresponding to the barycenter model (fully trust in $\mathcal{M}$) and the most distorted model (no trust in $\mathcal{M}$) which corresponds to the deep uncertainty case. In this work, we focus on the case that the investor is fully confident regarding the plausibility of the provided models and we refer to the other very interesting cases for future work in the subject.

\subsection{Optimal static hedging strategies under the model uncertainty framework}

Let us state in a mathematical formulation the static hedging problem that the investor faces. Consider that the investor has the obligation (as a charterer) to cover the cost of carrying for a cargo of size $D$ in tonnes (or resp. to travel $D$ days) at a future time $t=T$ which depends on the spot freight rate (which expresses either the freight service price per ton or per day) that will hold at this time in the physical market, $S(T)$. As a result, the obligation that has to be covered by the charterer is 
\begin{equation}\label{phi-1}
\Phi(S) = D S(T)
\end{equation}
or if an FFA contract has been conducted on the spot prices $S$ depending on some dates $T_1,T_2,...,T_N$, the obligation is expressed through the relation
\begin{equation}\label{phi-2}
\Phi(S) = D \frac{1}{N}\sum_{i=1}^N S(T_i).
\end{equation}

The second type of deals is preferred when both participants desire to avoid market volatility effects and therefore the freight service total price is determined through the arithmetic average of the spot prices at days $T_1,T_2,...,T_N$. Since the obligator is afraid of an uprising trend of the freight rates, he/she decides to built a hedging strategy to partly finance this forward contract. Due to the fact that the spot prices $S(t)$ are not traded, there is available to her/him a library of hedging instruments with underlying asset the route index $\Stil(t)$, i.e. European type call/put options, FFAs, etc. For simplicity, let us assume that the mapping referring to this library is
\begin{eqnarray}
{\bf \Psi}(\Stil) &=& D \begin{pmatrix} \Psi_1^{put}(\Stil) & \Psi_1^{call}(\Stil) & \Psi_2^{put}(\Stil) & \Psi_2^{call}(\Stil) \end{pmatrix}
\end{eqnarray}
where
\begin{eqnarray*}
	&&\Psi_1^{put}(\Stil) =  (K_1^{put} - \Stil(T))^+, \qquad \qquad \qquad
	\Psi_1^{call}(\Stil) =  (\Stil(T) - K_1^{call} )^+\\
	&&\Psi_2^{put}(\Stil) =  \left( K_2^{put} - \frac{1}{N}\sum_{i=1}^N\Stil(T_i)\right)^+, \qquad
	\Psi_2^{call}(\Stil) =  \left( \frac{1}{N}\sum_{i=1}^N\Stil(T_i) - K_2^{call} \right)^+
\end{eqnarray*}
containing both European put and call options alongside with FFA-type put and call options on the freight rate prices in the related route with strike prices $K_1^{put}, K_1^{call},K_2^{put},K_2^{call}$, respectively. The investor builds her/his hedging portfolio with respect to an allocation vector 
$${\bm \theta} = (\theta_1, \theta_2, \theta_3,\theta_4)' \in \Theta \subset \R^4$$ and the portfolio value at time $t=T$ will be
\begin{equation}
V_H(T;{\bm \theta}) = \langle {\bf \Psi}(\Stil,T), {\bm \theta} \rangle
\end{equation}
After assuming such a position, the remaining risk at maturity $T$, using a quadratic loss function, can be represented by 
\begin{equation}\label{qloss}
	L({\bm \theta}) := \frac{1}{2}\left( \Phi(S) - V_{H}(T;{\bm \theta}) \right)^2
\end{equation}
where $\Phi(S)$ is the obligation determined by \eqref{phi-1} or \eqref{phi-2}. The goal of the proposed static hedging mechanism, is to hedge the paying obligation $\Phi$ depending on $S$, with the portfolio $V_H$ returns  which is built on a collection of derivatives written on $\Stil$, in order to reduce the freight risk of the position in $\Phi$. The use of the quadratic loss function can be interpreted as a penalization for diverging from the exact hedge position and corresponds to cases where both super- or sub-hedging may lead to losses for the investor. Note that in place of $\Phi$ could be used and other type of derivatives that are preferred in similar operations, e.g. exchange options like Margrabe option \cite{margrabe1978value} or other type of exotic derivatives.

\subsubsection*{Step 1: Choosing a robust to uncertainty pricing measure}

From relation \eqref{qloss} is evident that the possible outcomes of the loss function depend on the probability model describing the evolution of the processes $S(t)$ and $\Stil(t)$, therefore the problem needs to restated in a stochastic formulation. In particular, the optimal (static) hedging strategy could be determined as the minimizer of the problem
\begin{equation}\label{SH-1}
\min_{\btheta \in \Theta} \E_Q[ L(\btheta) ] 
\end{equation} 
where $\Theta$ denotes the set of all admissible hedging strategies and $Q$ is the probability law describing the dynamics of $S, \Stil$. However, since there is uncertainty regarding the true model $Q$, and given the multitude of models $\mathcal{M}$ that the investor have been provided with, a robust version of the problem \eqref{SH-1} is considered with respect to the probability model that is selected. In particular, the investor has to solve the minimax problem
\begin{eqnarray}\label{static-robust}
\min_{ \btheta\in \Theta } \max_{ Q\in\mathcal{Q}_{\eta}} \mathbb{E}_{Q}[ L(\btheta) ] 
\end{eqnarray}
where $\mathcal{Q}_{\eta}$ denotes the set of plausible probability models according to the aversion preferences of the investor from the set of the provided models $\mathcal{M}$. The set $\mathcal{Q}_{\eta}$ can be expressed as (using the Wasserstein metric sense)
\begin{equation}
\mathcal{Q}_{\eta} := \left\{ Q \in \mathcal{P}(\Omega) \,\, : \,\, \sum_{i=1}^m w_i W_2^2(Q,Q_i) \leq \eta, \,\, \forall w\in\Delta^{m-1} \right\}
\end{equation}
where $\eta>0$ denotes the sensitivity parameter quantifying the aversion intensity and $w$ the weighting vector for the models in prior set $\mathcal{M}$. Setting the minimal value $\eta_*$ that can be attained for the Fr\'echet variance of a model $Q$ from the set $\mathcal{M}$, is equivalent to employ the Wasserstein barycenter of the set (as defined in \eqref{wbar}) as the discounting measure. Clearly, taking values $\eta > \eta_*$ will lead to deformations of the barycentric model providing even worse estimates for expected loss compared to the ones that can be derived by trusting only the information provided in $\mathcal{M}$. Problem \eqref{static-robust} essentially means that the investor chooses the hedging strategy so as to minimize the maximum expected loss over the set of possible probability laws for $L$, thus making a hedging decision under the worst-case scenario (since the robust version of the problem is expressed as a maximization problem of the expected loss with respect to the admissible set of probability models). This problem can also be realized as a game where nature plays against the decision maker (investor), where the first player (nature) chooses the model tha will produce the worst outcomes (losses) for the other side while the second player (investor) seeks to be protected against worst-case outcomes.

\subsubsection*{Step 2: Derivation of the optimal hedging strategy}

Given that the probability law that is chosen as the discounting measure for the financial decision is the barycenter of the set $\mathcal{M}$ (following the previous discussion), i.e. the model $Q_* := Q_*(w)$ as determined in \eqref{wbar} for a certain choice of $w\in\Delta^{m-1}$, the solution to the robust version of the static hedging problem
\begin{equation}\label{SH-2}
\min_{\btheta\in\Theta} \E_{Q_*}[ L(\btheta) ]
\end{equation}
is obtained in analytic form as a least-square estimate. It is quite straightforward to show that the objective function of problem \eqref{static-robust} admits the quadratic formulation
\begin{equation}
	J(\btheta) = \langle \btheta, {\bf C}_{\Psib} \btheta \rangle - 2\langle \btheta, C_{\Psib, \Phi} \rangle + C_{\Phi} 
\end{equation}
where
\begin{eqnarray}
	{\bf C}_{\Psib} &:=& \E_{Q_*}\left[ \Psib(\Stil)^T \Psib(\Stil) \right] \in \mathbb{P}(d)\\ 
	C_{\Psib,\Phi} &:=& \E_{Q_*}\left[  \Psib(\Stil)^T \Phi(S)  \right] \in \R^d\\
	C_{\Phi} &:=& \E_{Q_*}[ \Phi(S)^2 ] \in \R_+ .
\end{eqnarray}
Given that the set $\Theta$ is convex, the above objective function admits a unique solution which is obtained by taking first order conditions to  
\begin{equation}\label{sol-1}
	\btheta_* = {\bf Proj}_{\Theta} \left( C_{\Psib}^{-1} C_{\Psib,\Phi} \right)
\end{equation}
with the ${\bf Proj}_{\Theta}(\cdot)$ denoting the projection operator from $\R^2$ to the feasible set $\Theta$\footnote{ ${\bf Proj}_{\Theta}(x) := \arg \min_{y \in \Theta} \|y-x\|_2^2$ }.
Note that the calculation of the robust solution \eqref{sol-1} relies on our capability to produce samples from the pricing measure $Q_*$. Therefore it is a crucial step in the whole procedure the characterization of $Q_*$ or at least to have the ability to simulate samples from this model. For the case discussed in this paper (please see Section \ref{sec-4}) this task is straight-forward. However, for more complex models one has to be aware of the computational complexity that may face in obtaining $Q_*$. Possibly, in some cases, it would be a more profitable strategy to treat directly the minimax problem stated in \eqref{static-robust} in terms of required computational time.

Depending on the available hedging instruments in $\Psib(\Stil)$, problem \eqref{SH-2} might need to be enhanced with some regularization terms with respect to the hedging strategy $\btheta$ to simultaneously perform selection of the hedging instruments that are more effective (see e.g. sparse-type regularization schemes  \cite{corsaro2021fused, pun2021sparse}). For instance, hedging library $\Psib(\Stil)$ may contain both put and call options on various maturities, so employing penalties like LASSO, will provide hedging schemes that will completely omit budget allocation to financial instruments that cannot have a significant impact in reducing risk. In this direction, the penalized version of the problem might be re-written as
\begin{equation}\label{SH-3}
\min_{\btheta\in\Theta} \left\{ \E_{Q_*}[ L(\btheta) ] + R(\btheta) \right\}
\end{equation}
where $R(\btheta)$ denotes the employed regularization term (e.g. for LASSO regularization term we have $R(\btheta) = \lambda \| \btheta \|_1$ with $\lambda>0$ denoting the sensitivity parameter of the regularization).

\subsection{Determination of the distribution for the aggregate model}\label{sec-3.3}

A standard step before the selection of the optimal hedging strategy, is the determination of the (aggregate) model under which the optimal strategy (hedging ratios) will be derived. This model is obtained by combining the information provided through the prior set of models $\mathcal{M}$ and weighting them, according to the trust that the risk manager allocates to each information source. Setting these preferences through the determination of the weighting vector $w\in\Delta^{m-1}$, the aggregate model $Q_*$ is obtained through the solution of the problem stated in \eqref{wbar}. Attempting to provide some examples where closed-form solutions or at least semi-analytic expressions can be derived, we constrain ourselves to work with probability models that are members of some Location-Scatter family for which the aggregate model can be explicitly characterized by applying the equations \eqref{wloc} and \eqref{wcov}. As a result, the considered prior sets of models for the freight rates related to the hedging problem \eqref{SH-1}, are of the same type but with different parameters. The model settings that are considered are briefly presented below. 

\subsubsection*{Black's model (GBM)}

As a first example we assume the typical GBM model for both processes (spot freight rates and FFA rates) with correlated Brownian motions described by the dynamics: 
\begin{eqnarray}\label{mod-1}
	\left\{
	\begin{array}{lr}
		dS(t)    = \mu S(t) dt + \sigma S(t) dW(t) & \mbox{(spot dynamics)}\\
		d\Stil(t) = \tilde{\mu} \Stil(t) dt + \tilde{\sigma} \Stil(t) d\widetilde{W}(t) & \mbox{(FFA-rate dynamics)}\\
		\E[dW(t)d\widetilde{W}(t)] = \rho dt & 
	\end{array} \right.
\end{eqnarray}
Employing the logarithmic prices in both cases, the bivariate random variable 
$$X(t) := (X_1(t), X_2(t))' =  ( \log(S(t)), \log( \Stil(t) ) )'$$ 
is normally distributed for fixed $t$. In particular, for a fixed time instant $t$ and given the information collected up to time instant $s<t$, $\mathcal{F}_s$, the random variable $X(t)$ is distributed according to $N({\bm m}_ {t|s}, C_{t|s})$ where the location parameter vector is 
\begin{equation}
	{\bm m}_ {t|s} :=  
	\begin{pmatrix} 
		m_{t|s} \\ \tilde{m}_{t|s} 
	\end{pmatrix} = 
	\begin{pmatrix} 
		\E[ \log(S(t)) | \mathcal{F}_s ] \\ \E[ \log(\Stil(t)) | \mathcal{F}_s ]
	\end{pmatrix} = 
	\begin{pmatrix} 
		\log( S(s) ) + \left( \mu - \frac{\sigma^2}{2}\right)(t-s) \\ 
		\log( \Stil(s) ) + \left( \tilde{\mu} - \frac{\tilde{\sigma}^2}{2}\right)(t-s)
	\end{pmatrix}
\end{equation}
and the dispersion matrix is defined as 
\begin{equation}
	C_{t|s} := 
	\begin{pmatrix} 
		v_{t|s} & \gamma_{t|s}\\ 
		\gamma_{t|s} & \tilde{v}_{t|s} 
	\end{pmatrix} = 
	\begin{pmatrix} 
		\sigma^2 (t-s) & \rho \sigma \tilde{\sigma} (t-s)\\ 
		\rho \sigma \tilde{\sigma} (t-s) & \tilde{\sigma}^2 (t-s)
	\end{pmatrix}
\end{equation}
with $v_{t|s} := Var( \log(S(t)) | \mathcal{F}_s )$, $\tilde{v}_{t|s} := Var( \log(\Stil(t)) | \mathcal{F}_s )$ and $\gamma_{t|s} := Cov( \log(S(t)), \log(\Stil(t)) | \mathcal{F}_s )$.

\subsubsection*{Schwarz's model (MR)}

As a second interesting case, the MR model is assumed for both processes (spot freight rates and FFA rates) with correlated Brownian motions described by the dynamics: 
\begin{eqnarray}\label{mod-2}
	\left\{
	\begin{array}{lr}
		dS(t)    = \alpha (\mu - S(t)) dt + \sigma dW(t) & \mbox{(spot dynamics)}\\
		d\Stil(t) = \tilde{\alpha} (\tilde{\mu} - \Stil(t)) dt + \tilde{\sigma} d\widetilde{W}(t) & \mbox{(FFA-rate dynamics)}\\
		\E[dW(t)d\widetilde{W}(t)] = \rho dt & 
	\end{array} \right.
\end{eqnarray}
In this case, there is no need to perform some transformation to obtain an elliptical probability model. In fact, the bivariate random variable 
$$X(t) := (X_1(t), X_2(t))' =  ( S(t), \Stil(t) )'$$ 
is distributed for fixed $t$ according to the Gaussian $N({\bm m}_ {t|s}, C_{t|s})$ where the location parameter vector is determined by
\begin{equation}
	{\bm m}_ {t|s} :=  
	\begin{pmatrix} 
		m_{t|s} \\ \tilde{m}_{t|s} 
	\end{pmatrix} = 
	\begin{pmatrix} 
		\E[ S(t) | \mathcal{F}_s ] \\ \E[ \Stil(t) | \mathcal{F}_s ]
	\end{pmatrix} = 
	\begin{pmatrix} 
		e^{-\alpha (t-s)} S(s) + \mu \left( 1 -  e^{-\alpha (t-s)}\right) \\ 
		e^{-\tilde{\alpha} (t-s)}\Stil(s) + \tilde{\mu} \left( 1 -  e^{-\tilde{\alpha} (t-s)}\right)
	\end{pmatrix}
\end{equation}
and the dispersion matrix is 
\begin{equation}
	C_{t|s} := 
	\begin{pmatrix} 
		\frac{\sigma^2}{2\alpha}\left( 1 - e^{-2\alpha (t-s)} \right) & 
		\frac{\rho \sigma \tilde{\sigma}}{2} \sqrt{ \frac{ (1-e^{-2\alpha(t-s)})((1-e^{-2\tilde{\alpha}(t-s)}) }{\alpha \tilde{\alpha} } }    \\ 
		\frac{\rho \sigma \tilde{\sigma}}{2} \sqrt{ \frac{ (1-e^{-2\alpha(t-s)})((1-e^{-2\tilde{\alpha}(t-s)}) }{\alpha \tilde{\alpha} } }  &
		\frac{\tilde{\sigma}^2}{2\tilde{\alpha}}\left( 1 - e^{-2\tilde{\alpha} (t-s)} \right)
	\end{pmatrix}
\end{equation}
with $v_{t|s} := Var( S(t) | \mathcal{F}_s )$, $\tilde{v}_{t|s} := Var( \Stil(t) | \mathcal{F}_s )$ and $\gamma_{t|s} := Cov( S(t), \Stil(t) | \mathcal{F}_s )$.

\subsubsection*{Mixed Black-Schwarz model (GBM--MR)}

A mixed model version is also considered, where the spot freight rates are assumed to follow a GBM model, and the FFA-rates are assumed to follow the MR model with their stochastic dynamics described by
\begin{eqnarray}\label{mod-3}
	\left\{
	\begin{array}{lr}
		dS(t)    = \mu S(t) dt + \sigma S(t) dW(t) & \mbox{(spot dynamics)}\\
		d\Stil(t) = \tilde{\alpha} (\tilde{\mu} - \Stil(t)) dt + \tilde{\sigma} d\widetilde{W}(t) & \mbox{(FFA-rate dynamics)}\\
		\E[dW(t)d\widetilde{W}(t)] = \rho dt & 
	\end{array} \right.
\end{eqnarray}
In order to be able to characterize the probability model as a Location-Scatter family member, let us consider the random variable $X(t) = (X_1(t), X_2(t))' = (\log(S(t)), \Stil(t))'$ for any fixed $t$. In this case, the bivariate random variable $X(t)$ is normally distributed with location determined by the vector
\begin{equation}
	{\bm m}_ {t|s} :=  
	\begin{pmatrix} 
		m_{t|s} \\ \tilde{m}_{t|s} 
	\end{pmatrix} = 
	\begin{pmatrix} 
		\E[ \log(S(t)) | \mathcal{F}_s ] \\ \E[ \Stil(t) | \mathcal{F}_s ]
	\end{pmatrix} = 
	\begin{pmatrix} 
		\log( S(s) ) + \left( \mu - \frac{\sigma^2}{2}\right)(t-s) \\ 
		e^{-\tilde{\alpha} (t-s)}\Stil(s) + \tilde{\mu} \left( 1 -  e^{-\tilde{\alpha} (t-s)}\right)
	\end{pmatrix}
\end{equation}
while the respective dispersion matrix is determined as
\begin{equation}
	C_{t|s} = 
	\begin{pmatrix} 
		v_{t|s} & \gamma_{t|s}\\ \gamma_{t|s} & \tilde{v}_{t|s} 
	\end{pmatrix}
	=
	\begin{pmatrix}
		\sigma^2 (t-s) & \rho \sigma \tilde{\sigma} \sqrt{(t-s)\frac{ 1 - e^{-2\tilde{\alpha}(t-s)} }{ 2\tilde{\alpha} } }\\
		\rho \sigma \tilde{\sigma} \sqrt{(t-s)\frac{ 1 - e^{-2\tilde{\alpha}(t-s)} }{ 2\tilde{\alpha} } } &
		\frac{\tilde{\sigma}^2}{2\tilde{\alpha}}\left( 1 - e^{-2\tilde{\alpha} (t-s)} \right)
	\end{pmatrix}
\end{equation}
where $v_{t|s} := Var( \log(S(t)) | \mathcal{F}_s )$, $\tilde{v}_{t|s} := Var( \Stil(t) | \mathcal{F}_s )$ and $\gamma_{t|s} := Cov( \log(S(t)), \Stil(t) | \mathcal{F}_s )$.

In all the above cases, the aggregate model $Q_*$ is characterized in semi-closed form by applying the equations \eqref{wbar}-\eqref{wcov}. The obtained model is then used for the derivation of the optimal (static) hedging strategy, i.e. for the calculation of the hedging ratios displayed in \eqref{sol-1} which are expressed in terms of expectations with respect to the aggregate probability model $Q_*$.

\section{Numerical illustration in choosing robust hedging strategies for mitigating freight risk}\label{sec-4}

In this section, some numerical experiments are performed based on the method and models discussed in Section \ref{sec-3}. First, a synthetic-data experiment is performed based on standard models used in practice (GBM, MR) for different levels of heterogeneity within the set of priors in order to assess the robustness of the method with respect to the issue of model ambiguity. Then, a second experiment is performed using real data from two main sea trade routes for different vessels at vertain time periods from the last two decades, where shocks were introduced to the world economy (Global Financial Crisis (GFC), BRICS financial crisis and COVID-19 pandemics) causing significant distirbunces to the shipping market.

\subsection{A synthetic data experiment for assessing robustness property of the method}\label{sec-4.1}

Based on the standard Black's (GBM) and Schwarz's (MR) models discussed in Section \eqref{sec-3.3}, numerical experiments are conducted under prior sets of varying homogeneity to determine the best hedging strategy for covering the open position of the problem stated in \eqref{SH-1}. To perform a comparison, a true model is assumed for the bivariate random variable $(S, \Stil)$, which is known to the experimenter but unknown to the risk manager. The risk manager, on account of incomplete information, is provided with a number of plausible models describing the evolution of the processes $(S, \Stil)$, which in certain aspects may be diverging, comprising the prior set $\mathcal{M}$. The risk manager wishes to static hedge her/his future obligations that have to be covered at time $T=T_N$ following the robust static hedging approach to counter the uncertainty concerning the exact financial amount that has to be covered at time $T$.

For assessing the performance of the proposed method, the obtained hedging strategy is compared to the optimal hedging strategy under the true model (if the latter was known). To check for the robustness of the findings, a number of experiments are performed for each one of the two considered models. These experiments are divided in three groups depending on the heterogeneity among the models that the investor is provided with. These groups are refferred from now on as the three different scenarios of information heterogeneity: the scenario of (a) Low Heterogeneity (LH), (b) Medium Heterogeneity (MH) and (c) High Heterogeneity (HH), determined by the degree of discrepancy of the models in $\mathcal{M}$. For each experiment, $m=3,5,10$ and $30$ different models that constitute the prior set $\mathcal{M}$ are generated as perturbations of the true model (by generating perturbations on the true model's parameters). Specifically, the various scenarios for the prior set $\mathcal{M}$ are generated by choosing random values for the model parameters by adding a noise term $\xi$ simulated from an appropriate Uniform distribution $U(\alpha,\beta)$ where $\alpha,\beta$ are chosen according to the parameter scaling and the homogeneity scenario that is considered.

As a specific example we consider the hedging problem stated in \eqref{static-robust}, with initial prices for the spot freight rate $S(0)=45$ and the FFA freight rate $\Stil(0)=40$ (both prices are  expressed in USD either per day or ton depending on what the ammount $D$ represents). The relevant strike prices for the call and put options are determined as $K_1^{put} = K_2^{put} = \Stil(0) \,\cdot \, 80\% $ and $K_1^{call} = K_2^{call} = \Stil(0) \,\cdot \, 120\%$, respectively, given that they refer to a one year hedging horizon. For shorter maturies, these values are scaled with respect to the time interval considered. Three different investing horizons are considered for all models and homogeneity levels of information: 3, 6 and 12 months (in trading days). Simulations of model paths are performed using the typical step $dt=1/252$, corresponding to the daily trading frequency. The true model parameters for each considered case are illustrated in Table \ref{tab-0} 

\begin{table}[ht!]
	\centering
	\begin{tabular}{l|rrrrrrr}
		\hline\hline
		{\bf Model }                                  & $\mu$ & $\tilde{\mu}$ & $\alpha$ & $\tilde{\alpha}$ & $\sigma$ & $\tilde{\sigma}$ & $\rho$ \\
		\hline
		Black's model (GBM) \eqref{mod-1}                      & 0.45 & 0.35 &       - &        - &  0.65 & 0.50 & 0.75 \\
		Schwarz's model (MR) \eqref{mod-2}                   &    36.00 &    36.00 & 5.50 &  4.00 &     24.00 &      20 & 0.80 \\
		\hline\hline
	\end{tabular}
	\caption{True model parameters used in the simulated experiments}\label{tab-0}
\end{table}

For each experiment $B=100000$ paths are simulated according to the distributions characte\-rizations in Section \ref{sec-3.3} while each prior heterogeneity scenario has been repeated $L=500$ times in an attempt to reduce the chance of reporting extreme cases. The obtained results are compared to the optimal decision obtained by the true model. The performance of the method is assessed through the typical statistical error indices: Bias, Mean Absolute Error (MAE), Root Mean Squared Error (RMSE) and Efficiency Index (the proportion of the generated variance under the adopted model to the variance of the true model), calculated with respect to the optimal results obtained by the true model.

\begin{figure}[ht!]
	\centering
	\includegraphics[width=4in]{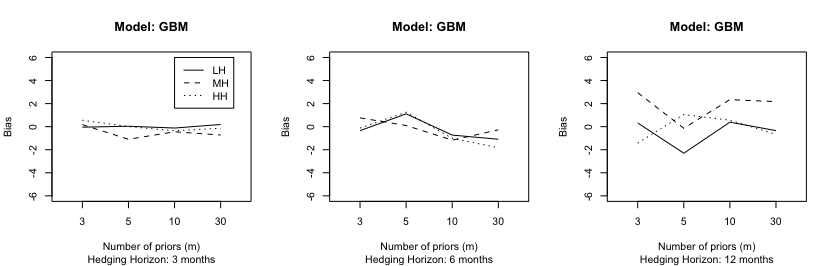}\\	
	\includegraphics[width=4in]{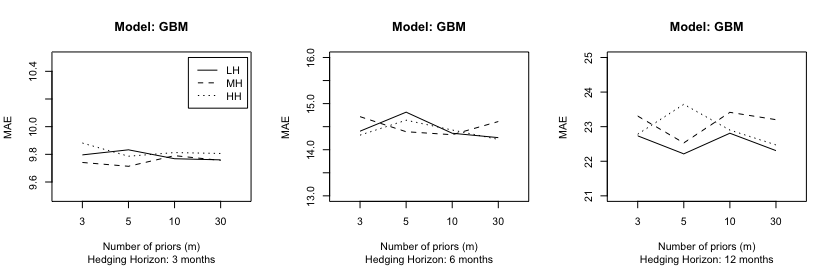}\\
	\includegraphics[width=4in]{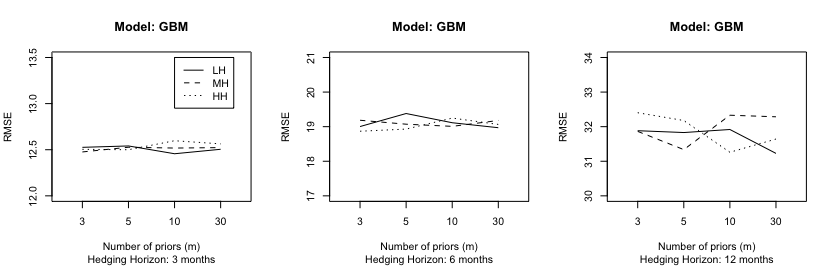}\\
	\caption{Error indices evolution for GBM model for different hedging horizons, prior set sizes and heterogeneity levels.}\label{fig-gbm-1}
\end{figure}

\begin{figure}[ht!]
	\centering
	\includegraphics[width=4in]{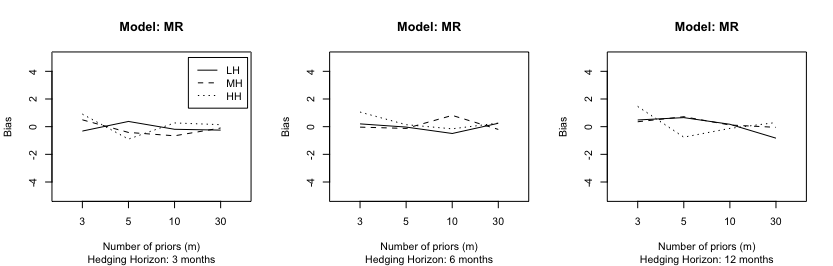}\\
	\includegraphics[width=4in]{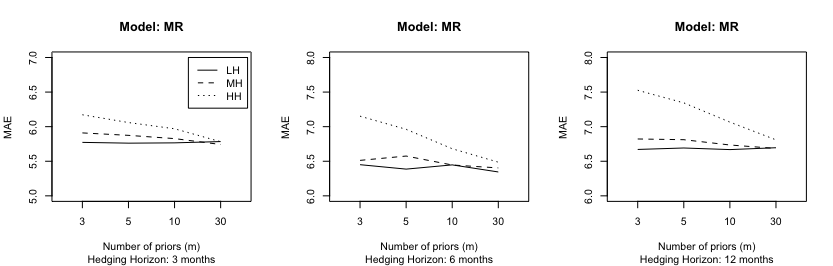}\\
	\includegraphics[width=4in]{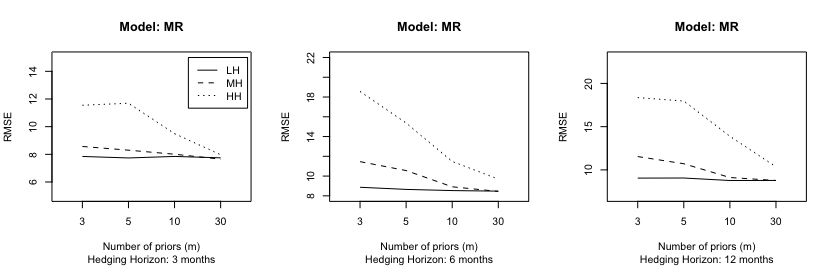}\\
	\caption{Error indices evolution for MR model for different hedging horizons, prior set sizes and heterogeneity levels.}\label{fig-mr-1}
\end{figure}

In Figure \ref{fig-gbm-1} and \ref{fig-mr-1} are displayed the error metrics obtained for each model setting under all heterogeneity scenarios, all prior sizes and for all investing horizons considered. For the GBM model, the three error indices (Bias, MAE and RMSE) seems to not present significant fluctuations around a certain level except the case of 12 months hedging horizon where a slightly increased variation is observed on account of the significant bigger time period that is considered. In this case the method displays a robust behaviour in the sense that the error magnitude remains roughly at the same levels regardless the heterogeneity level, the number of prior models and the length of the hedging horizon. For the MR model, bias remains at the same levels for all cases similarly to the behaviour of the GBM model. However, the magnitude of error (please see MAE and RMSE) it seems that is reduced as the number of priors grows for all hedging horizons, with more rapid rate for the MH level and less rapid for the HH level.

\begin{figure}[ht!]
	\centering
	\includegraphics[width=5in]{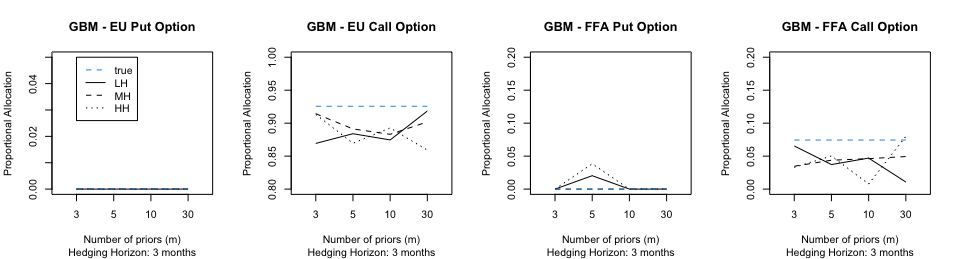}\\
	\includegraphics[width=5in]{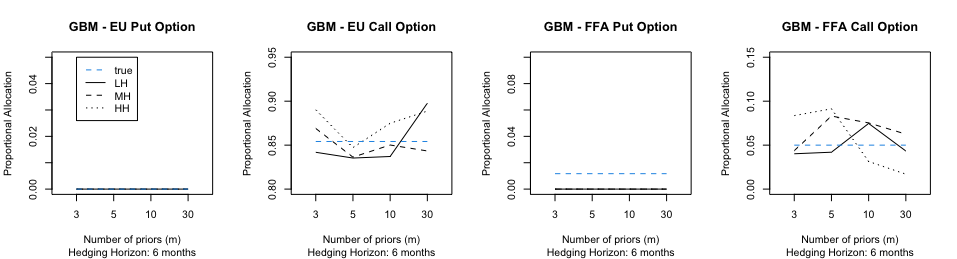}\\
	\includegraphics[width=5in]{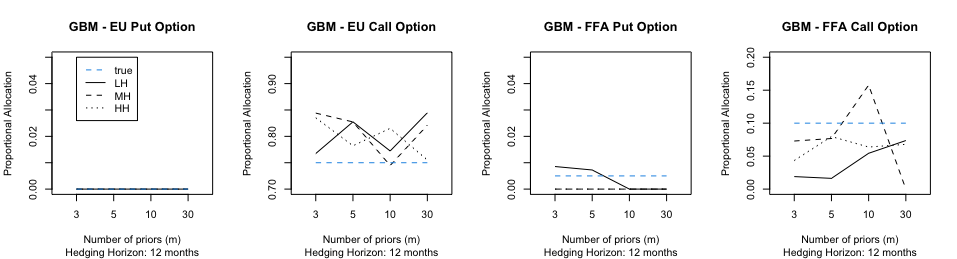}
	\caption{Approximation of true optimal hedging strategy under the GBM model}\label{fig-gbm-2}
\end{figure}

\begin{figure}[ht!]
	\centering
	\includegraphics[width=5in]{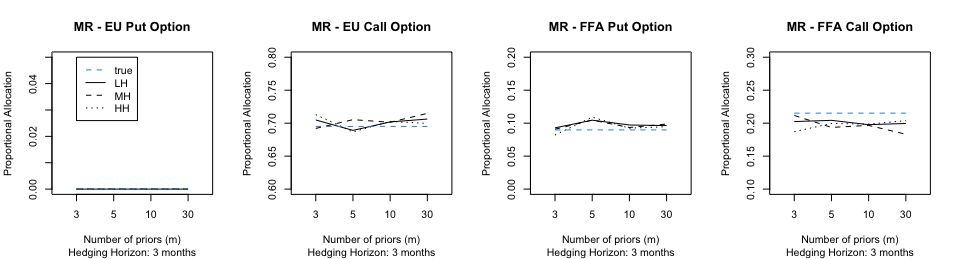}\\
	\includegraphics[width=5in]{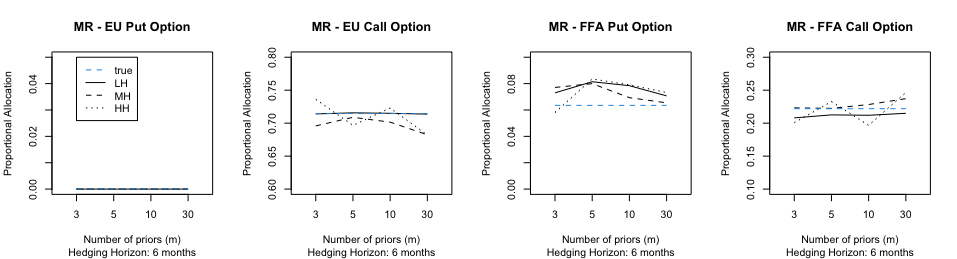}\\
	\includegraphics[width=5in]{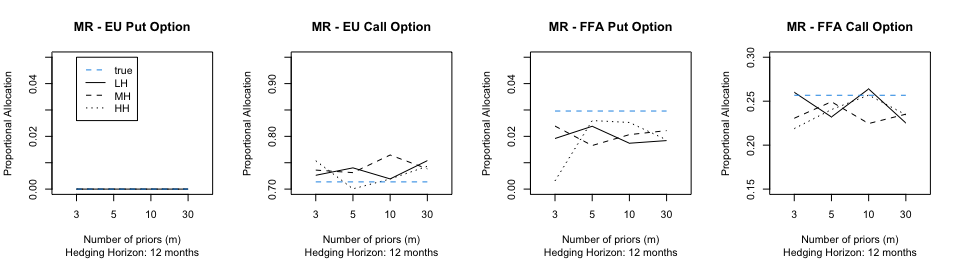}
	\caption{Approximation of true optimal hedging strategy under the MR model}\label{fig-mr-2}
\end{figure}

In Figures \ref{fig-gbm-2} and \ref{fig-mr-2} are displayed the true hedging strategy approximations for all assets in the hedging library and the three considered hedging horizons for both models. The nice behaviour in approximating the true situation for all heterogeneity levels in the prior set without significant differences either among the two models or the three hedging horizons. In both model settings considered (GBM and MR), the approximation error of the true hedging strategy seems to be at the same levels and not to be seriously affected by the heterogeneity of information in the prior set, indicating a robustness property of the method in recovering the true situation. Moreover, the error magnitude seems not to be significantly affected by the length of the hedging period. In general, the proposed method provides in most of the cases, estimates very close to the ones obtained by the true models, displaying quite a robust behaviour to the various information heterogeneity levels. 


\subsection{Case Study: Hedging freight risk at P1A and C2 routes}\label{sec-4.2}

As a second experiment, two main sea routes are considered and in particular routes P1A and C2 at different time spans of the last two decades where the world economy perturbed by significant shocks. The aforementioned routes concern voyages from Europe to the Americas through the Atlantic and different type of vessels, and in particular Panamax vessels (P1A route) and Capesize vessels (C2 route). The specifications for each trade route are provided in Table \ref{tab-4.2.1} while more information can be found in Baltic Exchange webpage\footnote{\scriptsize \url{https://www.balticexchange.com/en/data-services/routes.html}}.

\begin{table}[ht!]\scriptsize
	\centering 
	\begin{tabular}{  p{1cm}  |  p{1.2cm}  |  p{2.5cm}  |   p{9cm}   }
		\hline\hline
		{\bf Route} & {\bf Vessel } & {\bf Price Quotation} & {\bf Description}\\
		\hline
		P1A & Panamax & USD/Day & \scriptsize Dely Skaw-Gibraltar range, loading 15-20 days from the index date, for a transatlantic round voyage of 40-60 days, redelivery Skaw-Gibraltar range. 25\% weighting. 5.00\% total commission. Baltic Panamax vessel for Timecharter routes is a non-scrubber fitted vessel based on the following description: 82500mt dwt on 14.43m SSW draft, Max age 12 yrs, LOA 229m, beam 32.25m, TPC 70.5, 97000 cbm grain, 13.5 knots laden on 33mt fuel oil (380cs t) / 14 knots ballast on 31mt fuel oil (380cs t) + 0.1 MGO at sea, 11.5 knots laden on 22mt fuel oil (380cs t) / 12.5 knots ballast on 23mt fuel oil (380cs t) + 0.1 MGO at sea\\ 
		&		&	&	\\
		C2 & Capesize & USD/Ton & \scriptsize Tubarao to Rotterdam. 160,000lt iron ore, 10\% more or less in owner’s option, free in and out. 6 days, Sundays + holidays included all purposes. 6 hrs turn time at loading port, 6 hrs turn time at discharge port, 0.5\% in lieu of weighing. Freight based on long tons. Laydays/cancelling 20/30 days from index date. Age max 18 yrs. 5\% total commission.\\
		\hline\hline
	\end{tabular}
	\caption{Trade routes specifications as stated in Baltix Exchange webpage}\label{tab-4.2.1}
\end{table}

\begin{figure}[ht!]
	\begin{center}
	\includegraphics[width=3in]{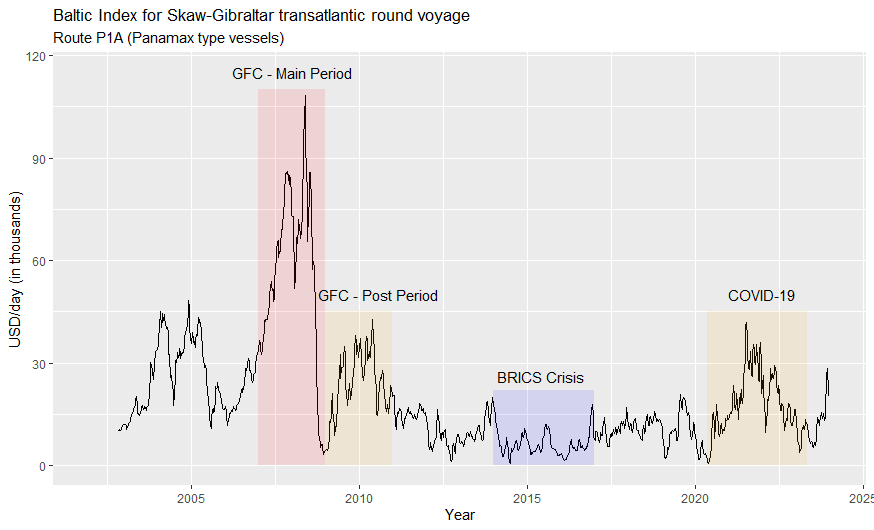}
	\includegraphics[width=3in]{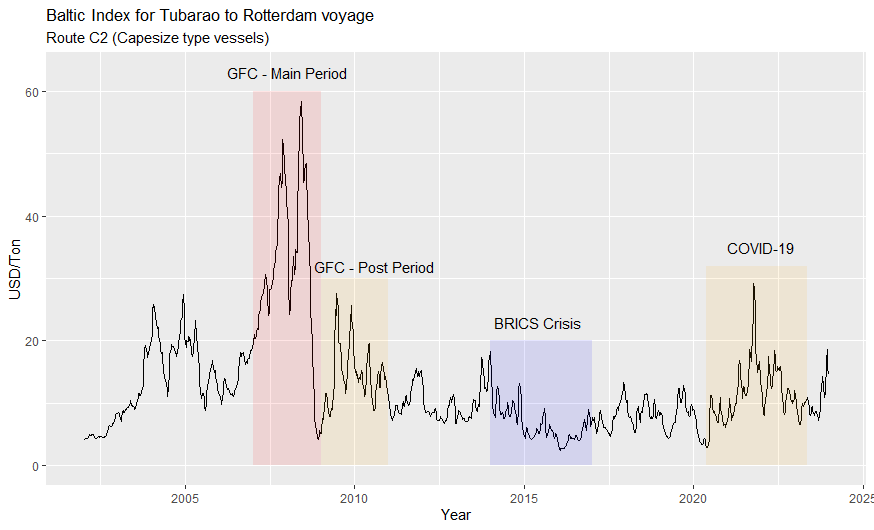}
	\end{center}
	\caption{Historical evolution for the time period 01/2002 - 12/2023 of the indices for two main routes for Panamax type vessels (P1A route - left figure) and for Capesize type vessels (C2 route - right figure). Data Source: Clarksons Research Portal}\label{fig-4.2.1}
\end{figure}

Three different case studies are considered in the last two decades for both routes related to events that led to significant disturbances of the world in economy. The Global Financial Crisis (GFC) is the first case study covering the period 2007-2010. Due to the different characteristics displayed a further division is performed to this case, distinguishing to (a) the GFC main period (2007-2008) and (b) the GFC post period (2009-2010). The second case study concerns the period 2014-2016 where several significant financial events occurred to major economies of the world belonging to BRICS\footnote{\scriptsize the intergovernmental organization comprising Brazil, Russia, India, China, South Africa, Egypt, Ethiopia, Iran, and the United Arab Emirates.} (devaluation of the Russian rouble in 2014, Brazilian political crisis that transformed to a severe financial crisis in 2014, the Chinese stock market turbulence in 2015-16, the commodity price shock at the second half of 2014, etc) unavoidably affecting the shipping industry. The last case study concerns the period of COVID-19 pandemics, including the years 2020 - 2022. The effects of the aforementioned events are observed in the freight rate prices for both routes which display very similar patterns. Clearly, GFC introduced a great shock to the rates both downwards and upwards, financial problems of BRICS countries seams that led to lower prices of the freight services while the COVID period led to an increase of the volatility of freight rates similar to the one displayed at the GFC post period (please see Figure \ref{fig-4.2.1}).

\begin{table}[ht!]
	\centering
	\begin{tabular}{clc}
		\hline\hline
		{\bf Case Study} & {\bf Major Event (Scenario)} & {\bf Period Covered}\\
		\hline
		I-a & Global Financial Crisis - Main Period & 2007-2008\\
		I-b & Global Financial Crisis - Post Period & 2009-2010\\
		II & BRICS Crisis 										& 2014-2016\\
		III & COVID-19 Pandemics							& 2020-2022\\
		\hline\hline
	\end{tabular}
	\caption{List of the case studies examined with the relevant time periods}\label{tab-4.2.2}
\end{table}

The experiment concerns the hedging task (from the side of the charterer) of freight contracts that reach to maturity in a total period of six months time (6m). To account for potential seasonality effects in the freight rates, for each year of interest (within the three considered time intervals 2007-2010, 2014-2016 and 2020-2022) each month is treated as a separate maturity date of the contract. Due to the occurring shocks in the world economy by the aforementioned events, it is not clear which is the best model to describe the forthcoming period. To simplify the model setup, the two standard models discussed in Section \ref{sec-3} (GBM and MR) are employed as plausible models (however this is not restrictive for the presented approach as any other model could be implemented) and the uncertainty issue refers to the ammount of recent empirical evidence that may contain important information for the future evolution of the freight rate prices. For the purposes of the experiment, four different time windows are considered (i.e. four prior models) adjusting each model parameters to the  information collected by the previous one, two, three and four years, comprising a set of short and long memory prior models. Note that the credibility of each model is not assessed, therefore each available model is equally taken into account. However, the addition of a filtering mechanism that may re-allocate the weights according to each model's credibility would be a very interesting direction for future work.

For both routes of interest, hedging portfolios consisting of EU-type options and FFA-type options on the index value of the route are constructed (both put and call options) in order to hedge the actual value of the service (according to the spot freight rate). The results of the method are assessed by simulating $B=100000$ trajectories for each case (i.e. each month of maturity in each one of the years of interest) according to the probability model as determined in Section \ref{sec-3.3}. The hedging performance of each portfolio is evaluated through the following metrics:

\begin{itemize} 
\item[(a)] \emph{Actual Loss Percentage}: the relative extra loss occurred at the maturity of the contract (at the end of the 6 months period) comparing to the value of the freight service at the time the contract is settled (at day 1 of the contract) and taking into account the contribution of hedging portfolio. Therefore, positive values are interpreted as proportional loss while negative values are interpreted as proportional profit.

\item[(b)] \emph{Hedging Effectiveness}: the ratio of the variance of the hedged loss to the variance of the unhedged loss. A standard index for measuring performance of hedging portfolios, values close to 1 indicate excellent hedging performance while values close to 0 indicate very poor performance.
\end{itemize}

\begin{figure}[ht!]
	\centering
	\includegraphics[width=4in]{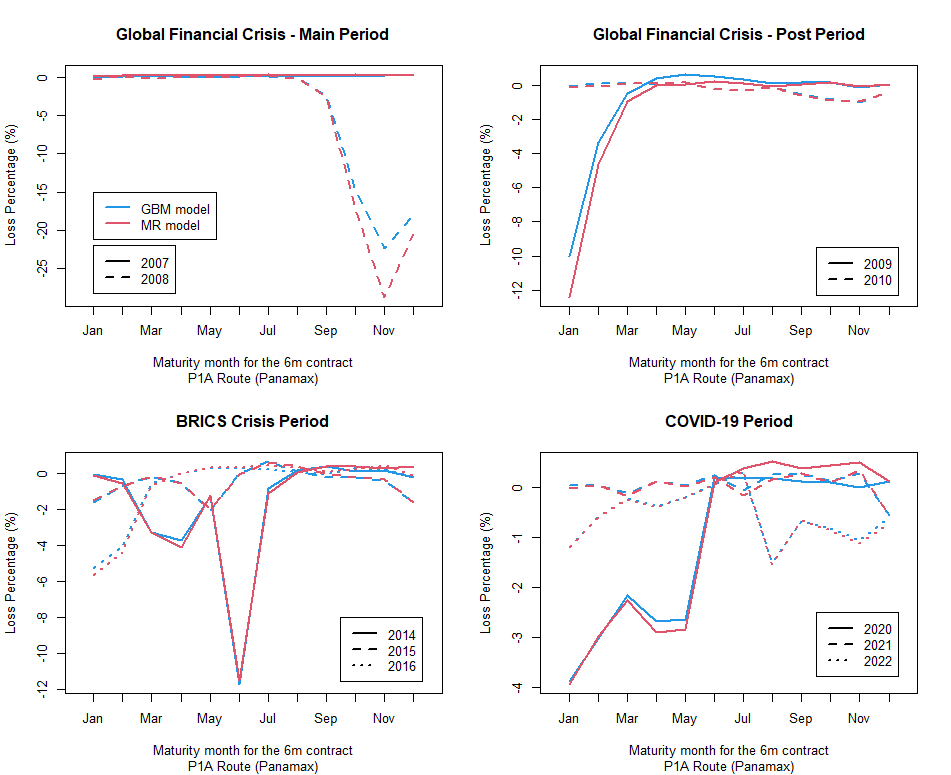}\\
	\includegraphics[width=4in]{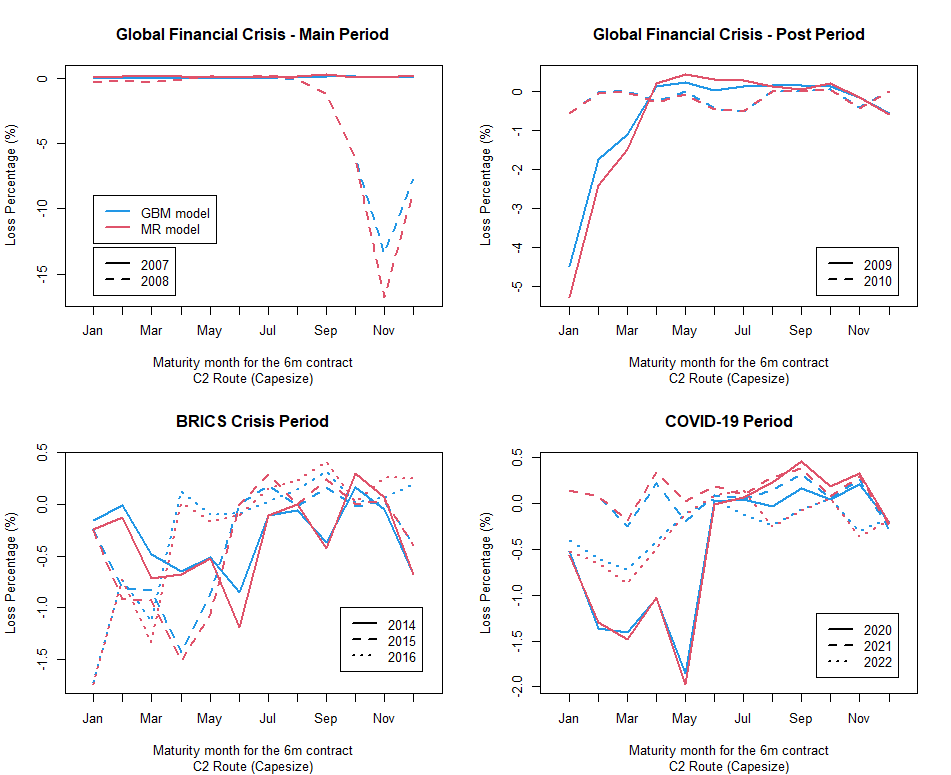}
	\caption{Actual Loss Percentage (per month of maturity)	of the risk hedging task in P1A (Panamax) and C2 (Capesize) routes for the considered time periods under the GBM and MR model considerations.}\label{fig-ALP}
\end{figure}

\begin{figure}[ht!]
	\centering
	\includegraphics[width=4in]{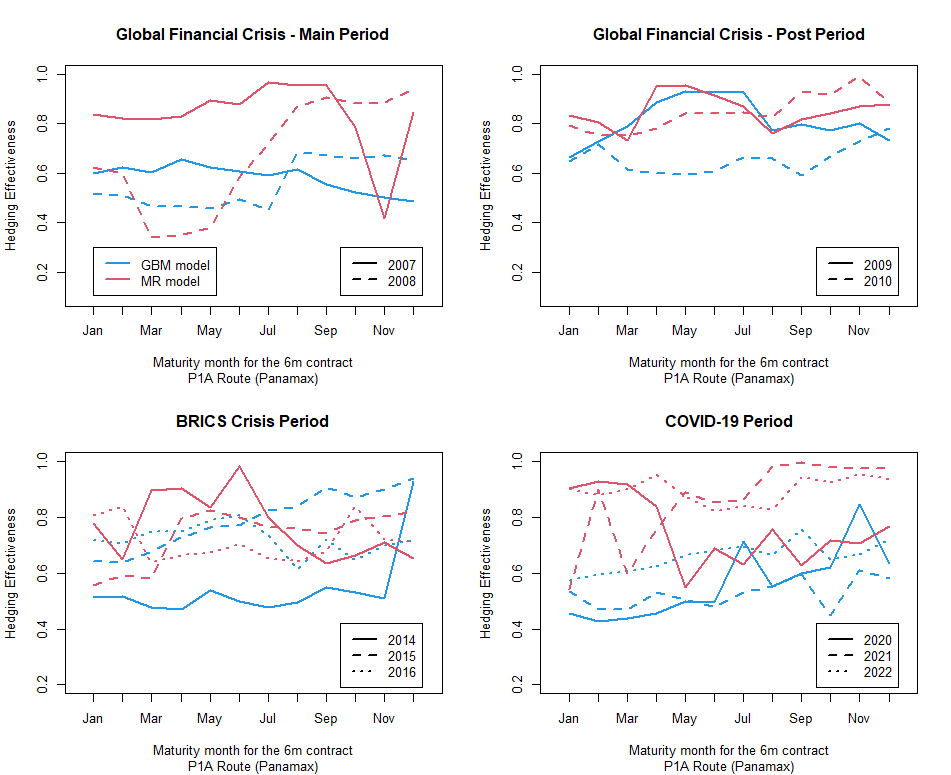}\\
	\includegraphics[width=4in]{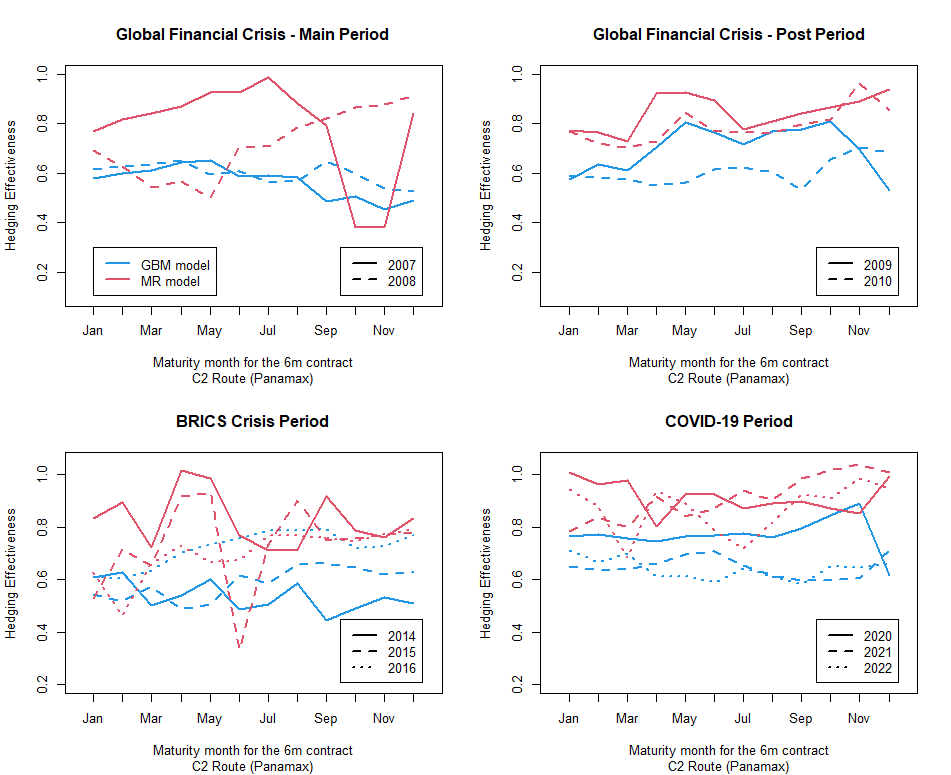}
	\caption{Hedging Efficiency (per month of maturity)	of the risk hedging task in P1A (Panamax) and C2 (Capesize) routes for the considered time periods under the GBM and MR model considerations.}\label{fig-HE}
\end{figure}

These metrics are illustrated in Figures \ref{fig-ALP} and \ref{fig-HE} for both routes and for all model considerati\-ons (GBM and MR model). Concerning the loss percentage, similar patterns are displayed for both routes at all scenarios and for both modelling approaches with small fluctuations. In particular, in GFC case extra profit is obtained at the last months of 2008 while extra profit is obtained at the first months of 2009 due to the rapid decrease and increase observed in the freight rates at these periods. During the BRICS crisis, a significant volatility is observed on the loss patterns among the various years with a significant profit to observed at June of 2014 (following the rapid derease in the freight prices this period). However, the profit of margin seems to be higher for operations related to Panamax vessels comparing to Capesize since the profit percentage is significant higher. At COVID-19 period, from May 2020 and afterwards the profit margin significantly reduced while at the peak period (at year 2021) are observed positive loss percentages with maximum value around 0.5\%.

In Figure \ref{fig-HE} is illustrated the hedging effectiveness of the method. In general, there is an empirical rule that characterizes as highly effective hedges that return values equal or greater than 0.8 to the index value. For the main period of GFC (2007-08), only the MR model led to highly effective hedging until November, 2007 where a major drop in the hedge performance. The nice behaviour of the model is recovered aroun August, 2008 and is maintained till the end of 2010. Note that for both routes the MR model displayed significanlty better performance at the biggest part. In the period of BRICS crisis, at year 2014 MR model is more preferable than GBM since led to better hedging performance while the following two years both models provide similar average hedging performances (around 0.60--0.80). At COVID period, the MR model provides better hedging effectiveness comparing to GBM for most months at all year, displaying an index value greater or equal than 0.80 with an exception on the first months of 2020 and 2021 for the Panamax route.  Finally, in Table \ref{tab-4.2.3} are displayed the optimal (static) hedging allocations derived for each available hedging instrument for all cases.

\begin{table}[ht] \tiny
	\centering
	\rotatebox{90}{
	\begin{tabular}{r|rrrr|rrrr|rrrr|rrrr} 
		\hline\hline
		&\multicolumn{16}{|c}{Contract Maturity (in quarters)}\\
		\hline
		Year & Q1 & Q2 & Q3 & Q4 & Q1 & Q2 & Q3 & Q4 & Q1 & Q2 & Q3 & Q4 & Q1 & Q2 & Q3 & Q4 \\ 
		\hline
		& \multicolumn{4}{c|}{EU put option} & \multicolumn{4}{c|}{EU call option} & \multicolumn{4}{c|}{FFA put option} & \multicolumn{4}{c}{FFA call option}\\		
		\hline
		& \multicolumn{16}{|c}{\mbox{}}\\
		&\multicolumn{16}{|c}{\it P1A Route - Geometric Brownian Motion (GBM) model} \\
		& \multicolumn{16}{|c}{\mbox{}}\\
		\hline
		2007 & 0.00 & 0.00 & 0.00 & 0.00 & 0.95 & 0.94 & 0.89 & 0.86 
				  & 0.01 & 0.00 & 0.00 & 0.00 & 0.04 & 0.13 & 0.12   & 0.14 \\
		2008 & 0.00 & 0.00 & 0.00 & 0.00 & 0.98 & 0.96 & 0.95 & 0.81 
				  & 0.00 & 0.00 & 0.08 & 0.09 & 0.02 & 0.04 & 0.18 & 0.28 \\   
		2009 & 0.00 & 0.00 & 0.00 & 0.00 	  & 0.72 & 0.37 & 0.51 & 0.69 & 0.02 & 0.00 & 0.00 & 0.02 & 0.07 & 0.00 & 0.02 & 0.09 \\  
		2010 & 0.00 & 0.00 & 0.00 & 0.00 	  & 0.66 & 0.69 & 0.66 & 0.64 & 0.03 & 0.11 & 0.05 & 0.08 & 0.09 & 0.19 & 0.17 & 0.14 \\  
		2014 & 0.00 & 0.00 & 0.00 & 0.00 	  & 0.91 & 0.85 & 0.78 & 0.98 & 0.03 & 0.07 & 0.00 & 0.01 & 0.04 & 0.11 & 0.00 & 0.10 \\ 
		2015 & 0.00 & 0.00 & 0.00 & 0.00 	  & 0.73 & 0.52 & 0.37 & 0.26 & 0.08 & 0.07 & 0.04 & 0.04 & 0.17 & 0.27 & 0.21 & 0.25 \\    
		2016 & 0.00 & 0.00 & 0.00 & 0.00 	  & 0.61 & 0.63 & 0.70 & 0.61 & 0.07 & 0.01 & 0.02 & 0.01 & 0.04 & 0.04 & 0.19 & 0.31 \\    
		2020 & 0.00 & 0.00 & 0.00 & 0.00	 & 0.94 & 0.88 & 0.78 & 0.73 & 0.02 & 0.02 & 0.08 & 0.07 & 0.02 & 0.09 & 0.24 & 0.22 \\  
		2021 & 0.00 & 0.00 & 0.00 & 0.00 	 & 0.82 & 0.68 & 0.44 & 0.44 & 0.08 & 0.10 & 0.22 & 0.27 & 0.10 & 0.15 & 0.30 & 0.48 \\  
		2022 & 0.00 & 0.00 & 0.00 & 0.00	 & 0.57 & 0.59 & 0.39 & 0.46 & 0.07 & 0.15 & 0.14 & 0.10 & 0.13 & 0.23 & 0.30 & 0.27 \\
		\hline
		& \multicolumn{16}{|c}{ \mbox{} }\\
		& \multicolumn{16}{|c}{\it P1A Route - Mean Reversion (MR) model}\\
		& \multicolumn{16}{|c}{ \mbox{} }\\
		\hline
		2007 & 0.00 & 0.00 & 0.00 & 0.00 &  0.43 & 0.42 & 0.43 & 0.45  
				  & 0.29 & 0.28 & 0.21 & 0.55 & 0.28 & 0.31 & 0.36 & 0.00 \\  
		2008 & 0.00 & 0.00 & 0.07 & 0.10 &  0.61 & 0.81 & 0.25 & 0.18 
				  & 0.39 & 0.17 & 0.44 & 0.47 & 0.00 & 0.02 & 0.24 & 0.25 \\ 
		2009 & 0.02 & 0.00 & 0.00 & 0.00 & 0.56 & 0.82 & 0.73 & 0.61 
				  & 0.34 & 0.14 & 0.21  & 0.15 & 0.23 & 0.62 & 0.28  & 0.24\\  
		2010 & 0.00 & 0.00 & 0.00 & 0.00 &  0.81 & 0.53 & 0.43 & 0.32 
				  & 0.16 & 0.23 & 0.26 & 0.24 & 0.21 & 0.24 & 0.32 & 0.45 \\  
		2014 & 0.00 & 0.00 & 0.00 & 0.00 & 0.66 & 0.67 & 0.38 & 0.35 
				  & 0.15 & 0.11 & 0.41 & 0.48 & 0.13 & 0.23 & 0.21 & 0.17 \\    
		2015 & 0.00 & 0.00 & 0.00 & 0.00 & 0.64 & 0.63 & 0.28 & 0.49 
				  & 0.21 & 0.13 & 0.54 & 0.31 & 0.14 & 0.23 & 0.18 & 0.20 \\
		2016 & 0.00 & 0.00 & 0.00 & 0.00 & 0.56 & 0.40 & 0.36 & 0.49 
				  & 0.21 & 0.41 & 0.48 & 0.27 & 0.23 & 0.19 & 0.16 & 0.24 \\ 
		2020 & 0.00 & 0.00 & 0.00 & 0.00 & 0.84 & 0.64 & 0.29 & 0.40 
				  & 0.04 & 0.19 & 0.58 & 0.45 & 0.12 & 0.17 & 0.13 & 0.15 \\  
		2021 & 0.00 & 0.00 & 0.00 & 0.00 & 0.89 & 0.67 & 0.60 & 0.67 
				  & 0.09 & 0.09 & 0.04 & 0.09 & 0.13 & 0.24 & 0.36 & 0.25 \\  
		2022 & 0.00 & 0.00 & 0.00 & 0.00 & 0.55 & 0.44 & 0.27 & 0.27  
				  & 0.16 & 0.25 & 0.42 & 0.28 & 0.29 & 0.32 & 0.31 & 0.45 \\ 
		\hline
		& \multicolumn{16}{|c}{\mbox{}}\\
		& \multicolumn{16}{|c}{\it C2 Route - Geometric Brownian Motion (GBM) model} \\
		& \multicolumn{16}{|c}{ \mbox{} }\\
		\hline
		
		2007 & 0.00 & 0.00 & 0.00 & 0.00 & 0.97 & 0.94 & 0.83 & 0.87 & 0.01 & 0.00 & 0.00 & 0.01 & 0.04 & 0.15 & 0.17 & 0.22 \\ 
		2008 & 0.00 & 0.00 & 0.00 & 0.00 & 0.96 & 0.92 & 0.87 & 0.82 & 0.12 & 0.18 & 0.03 & 0.11 & 0.11 & 0.24 & 0.17 & 0.29 \\  
		2009 & 0.00 & 0.00 & 0.00 & 0.00 & 0.88 & 0.98 & 0.86 & 0.86 & 0.00 & 0.00 & 0.00 & 0.03 & 0.06 & 0.03 & 0.01 & 0.10 \\ 
		2010 & 0.00 & 0.00 & 0.00 & 0.00 & 0.98 & 0.86 & 0.92 & 0.83 & 0.02 & 0.07 & 0.00 & 0.01 & 0.00 & 0.12 & 0.04 & 0.17 \\
		2014 & 0.00 & 0.00 & 0.00 & 0.00 & 0.95 & 0.90 & 0.82 & 0.87 & 0.02 & 0.00 & 0.01 & 0.00 & 0.05 & 0.11 & 0.14 & 0.14 \\ 
		2015 & 0.00 & 0.00 & 0.00 & 0.00 & 0.88 & 0.86 & 0.80 & 0.69 & 0.00 & 0.01 & 0.01 & 0.00 & 0.00 & 0.09 & 0.09 & 0.17 \\ 
		2016 & 0.00 & 0.00 & 0.00 & 0.00 & 0.88 & 0.68 & 0.67 & 0.76 & 0.04 & 0.00 & 0.00 & 0.03 & 0.09 & 0.23 & 0.04 & 0.14 \\ 
		2020 & 0.00 & 0.00 & 0.00 & 0.00 & 0.64 & 0.70 & 0.81 & 0.66 & 0.10 & 0.04 & 0.03 & 0.00 & 0.10 & 0.24 & 0.26 & 0.10 \\   
		2021 & 0.00 & 0.00 & 0.00 & 0.00 & 0.82 & 0.84 & 0.69 & 0.62 & 0.01 & 0.06 & 0.12 & 0.12 & 0.06 & 0.04 & 0.22 & 0.27 \\ 
		2022 & 0.00 & 0.00 & 0.00 & 0.00 & 0.76 & 0.72 & 0.62 & 0.53 & 0.07 & 0.02 & 0.10 & 0.02 & 0.06 & 0.07 & 0.26 & 0.17 \\ 
		\hline
		&\multicolumn{16}{|c}{\mbox{}}\\
		&\multicolumn{16}{|c}{\it C2 Route - Mean Reversion (MR) model} \\
		&\multicolumn{16}{|c}{\mbox{}}\\
		\hline		
		2007 & 0.00 & 0.00 & 0.00 & 0.00 & 0.47 & 0.45 & 0.39 & 0.76 & 
					  0.25 & 0.22 & 0.37 & 0.14 & 0.27 & 0.33 & 0.23 & 0.10 \\  
		2008 & 0.00 & 0.00 & 0.00 & 0.05 & 0.69 & 0.77 & 0.39 & 0.34 & 
					  0.26 & 0.21 & 0.36 & 0.41 & 0.05 & 0.02 & 0.25 & 0.21 \\ 
		2009 & 0.00 & 0.00 & 0.00 & 0.00 & 0.47 & 0.51 & 0.52 & 0.60 & 
						0.29 & 0.14 & 0.22 & 0.14 & 0.24 & 0.35 & 0.26 & 0.26 \\   
		2010 & 0.00 & 0.00 & 0.00 & 0.00 & 0.69 & 0.50 & 0.38 & 0.32 & 
					0.14 & 0.25 & 0.37 & 0.35 & 0.17 & 0.25 & 0.25 & 0.33 \\  
		2014 & 0.02 & 0.01 & 0.00 & 0.00 & 0.35 & 0.37 & 0.35 & 0.24 & 
					0.42 & 0.24 & 0.39 & 0.54 & 0.21 & 0.38 & 0.27 & 0.21 \\  
		2015 & 0.00 & 0.00 & 0.00 & 0.00 & 0.52 & 0.62 & 0.17 & 0.25 & 
					0.27 & 0.10 & 0.69 & 0.57 & 0.22 & 0.46 & 0.14 & 0.18 \\
		2016 & 0.00 & 0.00 & 0.00 & 0.00 & 0.60 & 0.36 & 0.21 & 0.32 & 0.23 & 0.45 & 0.64 & 0.46 & 0.17 & 			0.20 & 0.15 & 0.22 \\  
		2020 & 0.00 & 0.00 & 0.03 & 0.04 & 0.36 & 0.59 & 0.14 & 0.20 & 0.12 & 0.14 & 0.67 & 0.39 & 0.27 & 			0.21 & 0.16 & 0.37 \\ 
		2021 & 0.00 & 0.00 & 0.00 & 0.00 & 0.57 & 0.37 & 0.36 & 0.49 & 0.19 & 0.29 & 0.22 & 0.27 & 0.24 & 0.34 & 0.42 & 0.24 \\  
		2022 & 0.01 & 0.00 & 0.00 & 0.00 & 0.58 & 0.47 & 0.34 & 0.35 & 0.40 & 0.21 & 0.35 & 0.23 & 0.29 & 0.31 & 0.30 & 0.42 \\ 
		\hline\hline
		\end{tabular} }
		\caption{Optimal hedging allocation of the available hedging instruments under the GBM and MR models for the trade routes P1A (Panamax) and C2 (Capesize)}\label{tab-4.2.3}
	\end{table}

\section{Discussion - Concluding Remarks}

In this paper, the problem of hedging freight rate risk in shipping operations under model uncertainty was studied. A framework for treating robustly the model ambiguity issue has been proposed, for the case where a multitude of prior models are available, under which optimal hedging strategies are derived. A key intermediate step in the procedure is the characterization of the pricing measure in semi-closed form emplloying the notion of Wasserstein barycenter in the space of probability models. The aforementioned probability barycenter employed as the aggregate model succeeds in simultaneously (a) combining potential conflicting prior models to a single one and (b) treating robustly the incurred uncertainty. Although the proposed method is introduced and implemented in a static framework, the same approach can be extended to a multi-stage or a dynamic setting, as well. As a next step, more sophisticated aggregation schemes may be considered allowing for a proper averaging of the information provided by the avialable models taking into account each model's validity in the spirit of the schemes discussed in \cite{papayiannis2018learning, koundouri2024consensus}.

The numerical experiments conducted in this paper, employ standard models that are preferred in the freight risk management practice, and implemented within the multiple-prior model framework. The hedging performance of the proposed method and its ability to uncover the true situatiion was assessed under different scenarios and within periods where the shipping market was at shock. First, in the synthetic data experiment the robustness of the method in recovering the true situation was tested under varying degrees of heterogeneity in the prior set, differrent lengths of the hedging horizon and different prior set sizes. The results indicated that the method's accuracy is not significantly affected by the level of heterogeneity within the prior set. Moreover, under pure Gaussian model considerations (e.g. Schwarz's model) the error magnitude may decline faster as the size of the set grows for the case of less heterogeneous priors. However, better accuracy may be obtained in all cases by the integration of the method with appropriate weighting schemes as mentioned above. In the real data experiment, the method's suitability was tested under differrent extreme scenarios and different model considerations with quite interesting findings. In particular, the levels of hedging effectiveness are acceptable for the greatest part of the studied hedging periods, while in the most cases, extra profit from the hedging operation is obtained. In general, the method seems to be capable to support robust decision making in circumstances where the balance of the market is significantly perturbed even under standard modelling considerations.



\bibliographystyle{apacite} 
\bibliography{freight_rate_risk_hedging}

\end{document}